\documentclass[
twocolumn,amsmath,amssymb,amsfonts,preprintnumbers,aps,prd,
longbibliography,bibnotes,floatfix
]{revtex4-2}
\usepackage{CJKutf8}
\usepackage{amsmath,amssymb,mathtools}
\usepackage{braket}
\usepackage{graphicx}
\usepackage{orcidlink}
\usepackage{capt-of}
\usepackage{xcolor}
\usepackage{tikz}
\usetikzlibrary{decorations.markings,arrows.meta}
\usepackage{hyperref}
\hypersetup{colorlinks=true,linkcolor=blue!45!black,
  citecolor=blue!45!black,urlcolor=blue!45!black,
  pdftitle={A Simplified Born Rule for Cosmological Correlators},
  pdfauthor={Fan Zhu}}
\newif\ifauthornotes
\authornotestrue

\renewcommand{\ap}{\alpha'}
\newcommand{\tildd}[1]{\widetilde{#1}}
\newcommand{\dd}{\mathrm d}
\newcommand{\Rpos}{\mathbb R_{>0}}
\newcommand{\cF}{\mathcal F}
\newcommand{\cB}{\mathcal B}

\newcommand{\Born}{\operatorname{Born}}
\newcommand{\Newt}{\operatorname{Newt}}
\newcommand{\Facet}{\operatorname{Facet}}
\newcommand{\Cut}{\operatorname{Cut}}
\newcommand{\ST}{\operatorname{ST}}

\newcommand{\Conn}{\operatorname{Conn}_{\geq2}}
\newcommand{\Univ}{\operatorname{Univ}}
\newcommand{\ind}{\mathbf 1}

\newcommand{\str}{\mathrm{str}}
\newcommand{\evec}{\mathbf e}
\DeclareRobustCommand{\BornMinkowskiGraph}[2][2.65mm]{%
  \begin{tikzpicture}[x=#1,y=#1,baseline=-.25ex,
    line cap=round,line join=round,
    bmg/edge/.style={draw=black,line width=.40pt},
    bmg/vertex/.style={circle,fill=black,draw=none,inner sep=0pt,
      minimum size=1.10pt}]
    \path[use as bounding box] (-.08,-.08) rectangle (1.08,.94);
    \ifcase#2
      \draw[bmg/edge] (0,0)--(.5,.86)--(1,0)--cycle;
      \node[bmg/vertex] at (0,0) {}; \node[bmg/vertex] at (.5,.86) {};
      \node[bmg/vertex] at (1,0) {};
    \or
      \draw[bmg/edge] (0,0)--(.5,.86);
      \node[bmg/vertex] at (0,0) {}; \node[bmg/vertex] at (.5,.86) {};
    \or
      \draw[bmg/edge] (0,0)--(1,0);
      \node[bmg/vertex] at (0,0) {}; \node[bmg/vertex] at (1,0) {};
    \or
      \draw[bmg/edge] (.5,.86)--(1,0);
      \node[bmg/vertex] at (.5,.86) {}; \node[bmg/vertex] at (1,0) {};
    \fi
  \end{tikzpicture}%
}

\makeatletter
\renewcommand{\paragraph}{%
  \@startsection{paragraph}{4}{\parindent}{\z@}{-1em}%
    {\normalfont\normalsize\bfseries}%
}
\makeatother

\makeatletter
\let\@bibdataout@init\relax
\let\pre@bibdata\@empty
\AtEndDocument{%
  \if@filesw
    \immediate\write\@auxout{\string\citation{REVTEX42Control,apsrev42Control}}%
  \fi
}
\makeatother
\begin{document}
\title{A Simplified Born Rule for Cosmological Correlators}
\author{Fan Zhu(\begin{CJK*}{UTF8}{gbsn}朱凡\end{CJK*})$^{1,2}$
\!\orcidlink{0009-0007-2188-4393}}
\email{zhufan25@gscaep.ac.cn}
\affiliation{$^1$ Graduate School of China Academy of Engineering Physics,
No. 10 Xibeiwang East Road, Haidian District, Beijing 100193, P.R. China}
\affiliation{$^2$ New Cornerstone Laboratory, Institute of Theoretical Physics,
Chinese Academy of Sciences, Beijing 100190, China}

\begin{abstract}
Cosmological correlators are natural observables of the primordial
universe and are related to wavefunction coefficients by the Born rule.
For equal-time in-in correlators in flat spacetime, we prove that
all double-boundary poles cancel for arbitrary graphs, including
in theories with local derivative interactions.
These cancellations occur before integration over spatial loop
momenta and yield a \emph{simplified Born rule} in terms of what
we call embedded wavefunction coefficients.
The resulting expansion reveals which physical poles can occur
together while keeping the interaction dependence explicit.
For conformally coupled $\operatorname{Tr}(\phi^3)$ theory in flat
spacetime, the simplified Born rule takes the form of a uniformly
weighted forest-cut expansion.
From its pole combinatorics, we construct what we call the
\emph{Born polytope}, whose canonical function reproduces the
normalized correlator for arbitrary graphs after matching its facet
supports to the physical energies. The same geometry also yields a
stringy formula for correlators.
\end{abstract}
\maketitle
\setcounter{secnumdepth}{2}

\section{Introduction}

Cosmological correlation functions encode the interactions and particle
content of the early universe. Their perturbative computation traditionally
uses the Schwinger--Keldysh, or in-in, formalism
\cite{Schwinger:1960qe,Keldysh:1964ud,Maldacena:2002vr,Weinberg:2005vy}.
Methods inspired by scattering amplitudes
\cite{Britto:2004recursion,Britto:2005proof,Bern:1994unitarity} have led to
reconstructions of inflationary correlators from symmetries and
singularities \cite{Arkani-Hamed:2018kmz}, cosmological cutting rules
from unitarity \cite{Goodhew:2020optical,Melville:2021cutting}, and
partial-energy recursion relations constrained by locality
\cite{Jazayeri:2021locality}.

The Born rule reconstructs correlators from wavefunction
coefficients~\cite{Maldacena:2002vr}, but singularities
of individual coefficients can cancel in the resulting sum.
The cosmological tree theorem relates loop wavefunction diagrams
to cut trees and explains the cancellations between loop and cut
contributions underlying the cosmological KLN
theorem~\cite{AguiSalcedo:2023tree}.
For nonderivative scalar theories in flat spacetime, the full-time
representation of in-in correlators~\cite{Donath:2024utn} makes the
absence of double-boundary poles manifest for arbitrary
graphs~\cite{Arkani-Hamed:2025mce}.

Positive geometry encodes singularities and their compatibility in
geometric boundaries and their intersections
\cite{Arkani-Hamed:2017tmz}.
Cosmological polytopes realize rational scalar wavefunction
integrands as canonical forms~\cite{cosmologicalpolytopes},
while related constructions describe scalar correlators
\cite{Benincasa:2024leu,Arkani-Hamed:2024jbp,Figueiredo:2025daa}.
Stringy canonical forms provide integral deformations that recover
polytope canonical functions as $\ap\to0$~\cite{stringycanonicalforms}.
For in-in correlators, the zonotope construction applies to
arbitrary graphs~\cite{geometryinin}. Building on this geometry,
Ref.~\cite{tropicalstringyinin} gives a stringy representation for
general graphs, together with an alternative based on the graph
correlahedron, or $\mathcal G$-cubeahedron, for trees.
The cosmological $\delta$-shift further connects tree-level scalar
and NLSM wavefunctions and correlators in
flat spacetime and de Sitter~\cite{Figueiredo:2026cosmodelta}.

In this Letter, we construct a simplified Born expansion for
arbitrary graphs that is free of double-boundary poles and makes
the surviving physical-pole patterns explicit.
We first illustrate the cancellation mechanism for
nonderivative scalars, then extend the construction to derivative
theories. Finally, we describe the Born polytope associated with the
scalar expansion and illustrate its canonical function and stringy
integral.

\section{Wavefunction and the Born rule}
We review wavefunction coefficients for conformally coupled massless
scalars in flat spacetime and the Born rule for equal-time
in-in correlators.
\begin{figure}[htbp]
	\centering
	\setlength{\abovecaptionskip}{3pt}
	\input{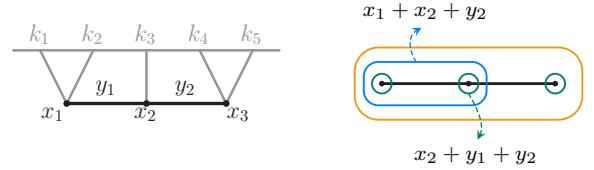}
	\caption{Left: the skeleton (black) of a five-point cubic Feynman
		diagram. Right: a maximal tubing of the same skeleton.}
	\label{fig:skeleton-graph}
\end{figure}

We encode each Feynman diagram
by a skeleton graph $G=(V,E)$, obtained by removing its external legs.
The site energy $x_v$ is the sum of the magnitudes of the external
momenta attached to $v$, and the edge energy $y_e$ is the magnitude of
the momentum carried by $e$. For example, in
Fig.~\ref{fig:skeleton-graph}, $x_1=|\mathbf k_1|+|\mathbf k_2|$ and
$y_1=|\mathbf k_1+\mathbf k_2|$.

For a connected vertex set $S\subseteq V$ or edge set $I\subseteq E$,
let $G[S]$ or $G[I]$ denote the corresponding vertex- or edge-induced
subgraph, respectively. We call such a connected subgraph a
\emph{tube}; two examples are labeled in Fig.~\ref{fig:skeleton-graph}.
The partial energy of a tube $K$ is
\begin{equation}
 X_K:=\sum_{v\in V(K)}p_v-2\sum_{e\in E(K)}y_e,
 \label{eq:induced-energy}
\end{equation}
where $p_v:=x_v+\sum_{v\in e\in E}y_e$ is the partial energy of a singleton.
All subgraphs inherit the singleton energies $p_v$ of the ambient graph $G$.
We also abbreviate $X_{G[S]}$ and $X_{G[I]}$ as $X_S$ and $X_I$,
respectively. 

For conformally coupled scalars, the wavefunction coefficient
associated with $G$ has the integral representation
\begin{equation}
 \Psi_G
 =\int_{\Rpos^{|V|}}\dd \mathbf{t}\,
 \prod_{v\in V}K_{x_v}(t_v)
 \prod_{e=uv\in E}G_{BB}(t_u,t_v;y_e).
 \label{eq:half-time-wavefunction}
\end{equation}
Here $\dd\mathbf{t}{:=}\prod_{v\in V}\dd t_v$, with each
Euclidean vertex time $t_v{>}0$. The scalar bulk-to-boundary
and bulk-to-bulk propagators are $K_x(t){:=}e^{-xt}$ and
$G_{BB}(t,t';y):=(e^{-y|t-t'|}-e^{-y(t+t')})/(2y)$,
respectively~\cite{Arkani-Hamed:2025mce}.
Using $(\partial_t+\partial_{t'})G_{BB}(t,t';y)=K_y(t)K_y(t')$,
integration by parts gives the recursive
formula~\cite{cosmologicalpolytopes}
\begin{equation}
 \Psi_H
 =\frac1{X_H}\sum_{e\in E(H)}
   \prod_{K\in\pi_0(H\setminus\{e\})}\Psi_K,
   ~~\Psi_{\{v\}}{=}\frac1{p_v},
 \label{eq:combinatorial-wavefunction-recursion}
\end{equation}
where $\pi_0(H)$ denotes the set of connected components of $H$.
Iterating this recursion
expresses $\Psi_H$ as a sum over
maximal tubings, namely maximal collections of pairwise nested
or vertex-disjoint tubes~\cite{Arkani-Hamed:2024jbp}, as illustrated
in Fig.~\ref{fig:skeleton-graph}.

For boundary field modes $\varphi_{\mathbf k}$, the Born rule gives
the equal-time $n$-point correlator~\cite{Maldacena:2002vr}:
\begin{equation}
 \langle\varphi_{\mathbf k_1}\cdots\varphi_{\mathbf k_n}\rangle
 =\frac{\int\mathcal D\varphi\,|\Psi[\varphi]|^2
 \varphi_{\mathbf k_1}\cdots\varphi_{\mathbf k_n}}
 {\int\mathcal D\varphi\,|\Psi[\varphi]|^2}.
 \label{eq:born-definition}
\end{equation}
Expanding the wavefunctional around its Gaussian part gives the
fixed-graph cut expansion~\cite{AguiSalcedo:2023tree,Goodhew:2020optical}
\begin{equation}
 \braket{G}=
 \sum_{I\subseteq E}
 \prod_{e\in I}\frac{1}{2\operatorname{Re}[\psi_2(y_e)]}
 \prod_{K\in\pi_0(G\setminus I)}2\operatorname{Re}(\Psi_K).
 \label{eq:ordinary-born}
\end{equation}
Here $\psi_2$ denotes the free-theory two-point wavefunction
coefficient, and $\braket G$ has its external free two-point
factors stripped off. Spatial loop
momenta are left unintegrated; state sums and index contractions are
implicit. Below we suppress $\operatorname{Re}$ on graph coefficients,
while retaining its action on the complete recursive expansion of
each original Born component.
For the canonically normalized massless scalars in flat spacetime
described by Eq.~\eqref{eq:half-time-wavefunction}, $\psi_2(y){=}y$ and
$\operatorname{Re}(\Psi_K){=}\Psi_K$.
Common scalar couplings are also omitted. 

The wavefunction coefficients in the Born expansion can contain
\emph{double-boundary poles}: a tube contains both endpoints of an
edge but not the edge itself, so its boundary crosses that edge twice.
They cancel in the correlator and are therefore
spurious~\cite{Donath:2024utn,Arkani-Hamed:2025mce}.

\section{The simplified Born rule}

\subsection{Conformally coupled Tr\texorpdfstring{$(\phi^3)$}{(phi3)} theory}
We illustrate the cancellation mechanism in
$\operatorname{Tr}(\phi^3)$ theory; the graph-level argument
applies to nonderivative scalar vertices of any valence.
We insert the recursion~\eqref{eq:combinatorial-wavefunction-recursion}
into the flat-space scalar specialization of
Eq.~\eqref{eq:ordinary-born}.
The cancellation illustrated in Fig.~\ref{fig:spurious-cancellation}
removes a double-boundary pole while preserving the cut-edge pole
and the remaining wavefunction factors.
\begin{figure}[htb]
	\centering
	\resizebox{0.85\linewidth}{!}{$\displaystyle\input{figures/phypole.tex}$}
	\caption{Cancellation of the double-boundary pole associated with the green
		tube $H=\{(12),(23)\}$ in $\braket{C_3}$. 
		Only the factors involved in the cancellation are shown.}
	\label{fig:spurious-cancellation}
\end{figure}
Repeating this procedure at successive stages of the recursion
eliminates all the double-boundary poles. Grouping the surviving terms
by the spanning tree in each component then gives the embedded
wavefunction coefficients and the uniformly weighted simplified
Born rule:
\begin{equation}
 \braket{G}=2^{|V|-|E|}
 \sum_{I\in\cF_{\mathrm{cut}}(G)}~
 \prod_{e\in I}\frac1{y_e}
 \prod_{T\in\pi_0(G\setminus I)}\Psi_T^{\mathrm{emb}},
 \label{eq:revised-born}
\end{equation}
where $\cF_{\mathrm{cut}}(G):=\{I\subseteq E:\ G\setminus I\text{ is a forest}\}$
is the set of forest cuts. 

The resulting \emph{embedded wavefunction coefficients} obey the
ordinary recursion~\eqref{eq:combinatorial-wavefunction-recursion},
with each tube energy $X_H$ replaced by the energy $X_{V(H)}$ of
the vertex-induced subgraph $G[V(H)]$:
\begin{equation}
 \Psi_H^{\mathrm{emb}}
 =\frac1{X_{V(H)}}\sum_{e\in E(H)}
   \prod_{K\in\pi_0(H\setminus\{e\})}\Psi_K^{\mathrm{emb}}\,.
 \label{eq:embedded-wavefunction}
\end{equation}
The singleton initial values remain $\Psi_{\{v\}}^{\mathrm{emb}}=1/p_v$.
The absence of double-boundary poles in correlators
\cite{Arkani-Hamed:2025mce} is thus already manifest
in each embedded wavefunction coefficient. For tree graphs, the two Born rules coincide.
Appendix~\ref{app:born-rule} illustrates the recursive
cancellation for the tailed bubble and gives the general proof.

We strip off the universal singleton factors and the uniform
coefficient by defining
\begin{equation}
	\tildd{\braket G}:=2^{|E|-|V|}\prod_{v\in V}p_v\,\braket{G}.
	\label{eq:normalized-correlator}
\end{equation}
\begin{widetext}
	\noindent\vbox{
		\centering
		\resizebox{0.9\textwidth}{!}{$\displaystyle
\begingroup
\newcommand{\CWSmallStep}{2.8mm}
\newcommand{\CWSmallLineWidth}{0.45pt}
\newcommand{\CWSmallDotRadius}{0.8pt}
\newcommand{\CWSingleDotStep}{\CWSmallStep}
\newcommand{\CWLargeStep}{7.5mm}
\newcommand{\CWLargeLineWidth}{0.8pt}
\newcommand{\CWLargeDotRadius}{1.5pt}
\newcommand{\CWArcHeight}{0.55}
\newcommand{\CWWaveSymbol}{\Psi}
\newcommand{\CWProductSpace}{\,}
\newcommand{\CWRowSpace}{1.2ex}
%
\tikzset{
  pics/cw/upper/.style={code={
    \draw (v1) .. controls (0.25,\CWArcHeight)
                       and (0.75,\CWArcHeight) .. (v2);
  }},
  pics/cw/lower/.style={code={
    \draw (v1) .. controls (0.25,-\CWArcHeight)
                       and (0.75,-\CWArcHeight) .. (v2);
  }},
  pics/cw/tail/.style={code={\draw (v2) -- (v3);}}
}
%
\newcommand{\CWSmallGraph}[2]{%
  \begin{tikzpicture}[
    x=\CWSmallStep,y=\CWSmallStep,baseline=-0.5ex,
    line width=\CWSmallLineWidth,line cap=round,line join=round]
    \coordinate (v1) at (0,0);
    \coordinate (v2) at (1,0);
    \coordinate (v3) at (2,0);
    \foreach \edge in {#1} {\pic {cw/\edge};}
    \foreach \vertex in {#2} {
      \fill (v\vertex) circle[radius=\CWSmallDotRadius];
    }
    \pgfmathsetmacro{\CWGraphLeft}{min(#2)-1-0.12}
    \pgfmathsetmacro{\CWGraphRight}{max(#2)-1+0.12}
    \pgfresetboundingbox
    \path[use as bounding box]
      (\CWGraphLeft,-0.60) rectangle (\CWGraphRight,0.60);
  \end{tikzpicture}%
}
\newcommand{\CWFullGraph}{\CWSmallGraph{upper,lower,tail}{1,2,3}}
\newcommand{\CWUpperTail}{\CWSmallGraph{upper,tail}{1,2,3}}
\newcommand{\CWLowerTail}{\CWSmallGraph{lower,tail}{1,2,3}}
\newcommand{\CWBubblePair}{\CWSmallGraph{upper,lower}{1,2}}
\newcommand{\CWUpperPair}{\CWSmallGraph{upper}{1,2}}
\newcommand{\CWLowerPair}{\CWSmallGraph{lower}{1,2}}
\newcommand{\CWTailPair}{\CWSmallGraph{tail}{2,3}}
\newcommand{\CWSingleVertex}[1]{%
  \begin{tikzpicture}[x=\CWSingleDotStep,y=\CWSmallStep,baseline=-0.5ex]
    \path[use as bounding box] (-0.12,-0.60) rectangle (2.12,0.60);
    \fill ({#1-1},0) circle[radius=\CWSmallDotRadius];
  \end{tikzpicture}%
}
\newcommand{\CWVertexOne}{\CWSingleVertex{1}}
\newcommand{\CWVertexTwo}{\CWSingleVertex{2}}
\newcommand{\CWVertexThree}{\CWSingleVertex{3}}
%
\newcommand{\CWWF}[1]{\CWWaveSymbol_{#1}}
%
\newcommand{\CWCorrelatorGraph}{%
  \begin{tikzpicture}[
    x=\CWLargeStep,y=\CWLargeStep,baseline=-0.5ex,
    line width=\CWLargeLineWidth,line cap=round,line join=round,
    every node/.style={font=\small,inner sep=1.4pt}]
    \coordinate (v1) at (0,0);
    \coordinate (v2) at (1,0);
    \coordinate (v3) at (2,0);
    \pic {cw/upper};
    \pic {cw/lower};
    \pic {cw/tail};
    \foreach \vertex in {1,2,3} {
      \fill (v\vertex) circle[radius=\CWLargeDotRadius];
    }
    \node[left]        at (v1) {$x_1$};
    \node[below right] at (v2) {$x_2$};
    \node[below]       at (v3) {$x_3$};
    \node[above] at (0.5,{0.75*\CWArcHeight})  {$y_2$};
    \node[below] at (0.5,{-0.75*\CWArcHeight}) {$y_1$};
    \node[above] at (1.5,0) {$y_3$};
  \end{tikzpicture}%
}
\newcommand{\CWRWF}[1]{\CWWaveSymbol_{#1}^{\mathrm{emb}}}
\left\langle\CWCorrelatorGraph\right\rangle
=\left\{\begin{array}{@{}l@{\quad}l@{}}
\text{Standard Born rule:}
&\begin{aligned}
&2\,\CWWF{\CWFullGraph}
 +\frac{1}{y_1}\,\CWWF{\CWUpperTail}
 +\frac{1}{y_2}\,\CWWF{\CWLowerTail}
 +\frac{2}{y_3}\,\CWWF{\CWBubblePair}\,\CWWF{\CWVertexThree}
 +\frac{1}{y_1y_2}\,\CWWF{\CWVertexOne}\,\CWWF{\CWTailPair}
 \\[1.2ex]
 &\quad+\frac{1}{y_1y_3}\,\CWWF{\CWUpperPair}\,\CWWF{\CWVertexThree}
 +\frac{1}{y_2y_3}\,\CWWF{\CWLowerPair}\,\CWWF{\CWVertexThree}
 +\frac{1}{y_1y_2y_3}\,\CWWF{\CWVertexOne}\,\CWWF{\CWVertexTwo}\,\CWWF{\CWVertexThree}
\end{aligned}
\\\noalign{\vskip 3ex}
\text{Simplified Born rule:}
&\begin{aligned}
&2^{3-3}\Biggl(
 \frac{1}{y_1}\,\CWRWF{\CWUpperTail}
 +\frac{1}{y_2}\,\CWRWF{\CWLowerTail}
 +\frac{1}{y_1y_2}\,\CWRWF{\CWVertexOne}\,\CWRWF{\CWTailPair}
 +\frac{1}{y_1y_3}\,\CWRWF{\CWUpperPair}\,\CWRWF{\CWVertexThree}
 \\[1.2ex]
 &\quad+\frac{1}{y_2y_3}\,\CWRWF{\CWLowerPair}\,\CWRWF{\CWVertexThree}
 +\frac{1}{y_1y_2y_3}\,\CWRWF{\CWVertexOne}\,\CWRWF{\CWVertexTwo}\,\CWRWF{\CWVertexThree}
 \Biggr)
\end{aligned}
\end{array}\right.
\endgroup%
$}
		\captionof{figure}{Comparison of the standard and simplified Born rules for the
			tailed-bubble graph in a flat-space theory of conformally coupled
			massless scalars with nonderivative interactions. The standard rule has
			$2^{|E|}=8$ cut sectors; the two retaining the bubble carry an extra factor
			of two. The simplified rule has six forest-cut sectors with the uniform weight $2^{3-3}=1$. }
		\label{fig:born-comparison}
	}
\end{widetext}
For the tailed bubble in Fig.~\ref{fig:born-comparison}, the simplified
rule gives
\begin{align}
	\tildd{\braket G}={}&
	\left(\frac1{y_1}+\frac1{y_2}\right)
	\left(
	\frac1{X_{123}X_{12}}+\frac1{X_{123}X_{23}}
	+\frac1{y_3X_{12}}
	\right)
	\nonumber\\[-1pt]
	&+\frac1{y_1y_2X_{23}}+\frac1{y_1y_2 y_3}.
	\label{eq:tailed-bubble}
\end{align}
Here $X_{12}$ denotes $X_{\{v_1,v_2\}}$, and similarly for $X_{23}$ and $X_{123}$.
Distinct subgraphs can have the same embedded wavefunction coefficient,
as illustrated by the two trees obtained by cutting either parallel edge
in this example.
The simplified rule makes the compatibility of $y_1$ and $X_{12}$
explicit, whereas it is obscured in the standard Born rule.
At a generic point of their intersection,
\begin{equation}
 \operatorname*{Res}_{X_{12}=0}
 \operatorname*{Res}_{y_1=0}\,
 \tildd{\braket G}
 =\left.\left(\frac1{X_{123}}+\frac1{y_3}\right)
   \right|_{y_1=X_{12}=0}\neq0.
 \label{eq:tailed-bubble-residue}
\end{equation}
The pole compatibility conditions in the simplified rule are
given in Eq.~\eqref{eq:facet-adjacency}.

\subsection{Derivative theory}\label{sec:derivative-extension}

We consider theories whose vertex and propagator
structures admit a differential-operator representation on scalar
kernels. For a fixed graph $G$, we encode these structures in
$\mathcal N_V^\partial$ and $\mathcal N_E^\partial$, respectively,
and write
\begin{equation}
 \begin{aligned}
 \Psi_G^\partial
 ={}&\int_{\Rpos^{|V|}}\dd\mathbf t\,
 \mathcal N_V^\partial\Biggl[
 \left(\prod_{i\in\mathrm{ext}}K_{k_i}(t_{v(i)})\right)\\
 &\qquad\times\mathcal N_E^\partial
 \left[\prod_{e=uv\in E}G_{BB}(t_u,t_v;y_e)\right]
 \Biggr].
 \end{aligned}
 \label{eq:derivative-half-time-wavefunction}
\end{equation}
Here $\mathrm{ext}$ denotes the external legs, and $v(i)\in V$
is the vertex to which leg $i$ attaches. Each leg has energy
$k_i=|\mathbf k_i|$, so $x_v=\sum_{i:v(i)=v}k_i$.
The operators
factor into vertex and edge operators with finite differential
expansions:
\begin{equation}
 \left\{
 \begin{aligned}
 &\mathcal N_V^\partial=\prod_{v\in V}\mathcal V_v,\quad\mathcal V_v=\sum_{\{r_{v,\ell}\geq0\}}C^{(v)}_{\{r_{v,\ell}\}}
 \prod_{\ell\ni v}
 \bigl(\partial_{t_v}^{(\ell)}\bigr)^{r_{v,\ell}},\\[2pt]
 &\mathcal N_E^\partial=\prod_{e=uv\in E}\Pi_e,\quad\Pi_{e}=\sum_{r,s\geq0}C^{(e)}_{rs}
 \bigl(\partial_{t_u}^{(e)}\bigr)^r
 \bigl(\partial_{t_v}^{(e)}\bigr)^s.
 \end{aligned}
 \right.
 \label{eq:graph-operator-products}
\end{equation}
For each vertex $v$, $\ell$ labels an incident internal or external
leg, and $r_{v,\ell}$ is the derivative order on that leg.
The operator $\mathcal V_v$ acts on the external modes and edge
kernels after $\mathcal N_E^\partial$ has acted, with each
derivative restricted to its indicated leg. Its coefficients
$C^{(v)}_{\{r_{v,\ell}\}}$ depend on the incident spatial momenta
and encode the local couplings and vertex tensors prescribed by
the mixed time--momentum Feynman rules~\cite{Melville:2021cutting}.
For each internal edge $e{=}uv$, $r$ and $s$ specify the derivative
orders at $u$ and $v$, respectively. The same operator $\Pi_e$
acts on $G_{BB}(t_u,t_v;y_e)$ for an uncut edge and on
$K_{y_e}(t_u)K_{y_e}(t_v)$ for a cut edge. Its coefficients
$C^{(e)}_{rs}$ depend on the edge momentum and encode the chosen
numerator tensors and boundary projectors.
All index contractions, matrix orderings, fermionic signs, and
state sums are inherited from the original graph. 

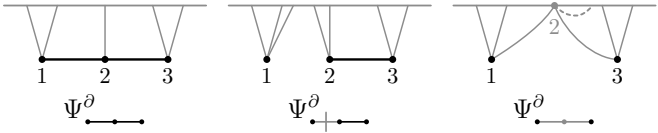
\begin{figure}[!htb]
 \centering
 \resizebox{\linewidth}{!}{$\displaystyle
\begingroup
\newcommand{\WFCUnit}{7mm}
\newcommand{\WFCColumnStep}{5.00}
\newcommand{\WFCBoundaryHeight}{1.20}
\newcommand{\WFCBoundaryHalfWidth}{2.25}
\newcommand{\WFCVertexStep}{1.40}
\newcommand{\WFCLegSpread}{0.34}
\newcommand{\WFCFormulaY}{-1.30}
\newcommand{\WFCFormulaFont}{\large}
\newcommand{\WFCVertexFont}{\normalsize}
\newcommand{\WFCBoundaryWidth}{0.80pt}
\newcommand{\WFCExternalWidth}{0.65pt}
\newcommand{\WFCInternalWidth}{1.06pt}
\newcommand{\WFCDotRadius}{1.55pt}
\newcommand{\WFCGrayTone}{45}
\newcommand{\WFCBoundaryLegSpan}{0.84}
\newcommand{\WFCBoundaryLegDepth}{0.30}
\newcommand{\WFCBoundaryLegWidth}{0.85pt}
\newcommand{\WFCBoundaryLegGray}{45}
\newcommand{\WFCBoundaryDashLength}{1.60pt}
\newcommand{\WFCBoundaryDashGap}{1.70pt}
\newcommand{\WFCCutCenterX}{-0.58}
\newcommand{\WFCCutHalfGap}{0.23}
\newcommand{\WFCSubgraphStep}{4.20mm}
\newcommand{\WFCSubgraphDotRadius}{1.10pt}
\newcommand{\WFCSubgraphLineWidth}{0.75pt}
\newcommand{\WFCSubgraphCutHalfLength}{0.35}
\newcommand{\WFCWave}{\Psi}
\newcommand{\WFCSup}{\partial}
\tikzset{
  wfc/boundary/.style={draw=black!\WFCGrayTone,line width=\WFCBoundaryWidth},
  wfc/external/.style={draw=black!\WFCGrayTone,line width=\WFCExternalWidth},
  wfc/boundary-leg/.style={draw=black!\WFCBoundaryLegGray,
    line width=\WFCBoundaryLegWidth,
    dash pattern=on \WFCBoundaryDashLength off \WFCBoundaryDashGap},
  wfc/internal/.style={draw=black,line width=\WFCInternalWidth},
  wfc/formula/.style={font=\WFCFormulaFont,anchor=base,inner sep=0pt},
  wfc/label/.style={font=\WFCVertexFont,inner sep=1pt}
}
%
\newcommand{\WFCPathSubscript}[4]{%
  \begin{tikzpicture}[x=\WFCSubgraphStep,y=\WFCSubgraphStep,
    baseline=-0.5ex,line cap=round,line join=round]
    \path[use as bounding box]
      (-0.16,{-max(0.23,\WFCSubgraphCutHalfLength+0.05)})
      rectangle (2.16,{max(0.23,\WFCSubgraphCutHalfLength+0.05)});
    \draw[draw=#1,line width=\WFCSubgraphLineWidth] (0,0)--(1,0);
    \draw[draw=#2,line width=\WFCSubgraphLineWidth] (1,0)--(2,0);
    \ifnum#4=1\relax
      \draw[draw=#1,line width=\WFCSubgraphLineWidth]
        (0.5,-\WFCSubgraphCutHalfLength)--(0.5,\WFCSubgraphCutHalfLength);
    \fi
    \fill[black] (0,0) circle[radius=\WFCSubgraphDotRadius];
    \fill[#3] (1,0) circle[radius=\WFCSubgraphDotRadius];
    \fill[black] (2,0) circle[radius=\WFCSubgraphDotRadius];
  \end{tikzpicture}%
}
%
\newcommand{\WFCOriginalFormula}{%
  \WFCWave^{\WFCSup}_{\WFCPathSubscript{black}{black}{black}{0}}}
\newcommand{\WFCVertexCutFormula}{%
  \WFCWave^{\WFCSup}_{\WFCPathSubscript{black!\WFCGrayTone}{black!\WFCGrayTone}{black!\WFCGrayTone}{0}}}
\newcommand{\WFCEdgeCutFormula}{%
  \WFCWave^{\WFCSup}_{\WFCPathSubscript{black!\WFCGrayTone}{black}{black}{1}}}
%
\newcommand{\WFCBoundary}{%
  \draw[wfc/boundary]
    (-\WFCBoundaryHalfWidth,\WFCBoundaryHeight)
    --(\WFCBoundaryHalfWidth,\WFCBoundaryHeight);
}
\newcommand{\WFCOuterLegs}{%
  \foreach \WFCX in {-\WFCVertexStep,\WFCVertexStep}{%
    \draw[wfc/external] (\WFCX,0)
      --({\WFCX-\WFCLegSpread},\WFCBoundaryHeight);
    \draw[wfc/external] (\WFCX,0)
      --({\WFCX+\WFCLegSpread},\WFCBoundaryHeight);
  }
}
\newcommand{\WFCMiddleLeg}{%
  \draw[wfc/external] (0,0)--(0,\WFCBoundaryHeight);
}
\newcommand{\WFCVertex}[2]{%
  \fill[black] (#1,0) circle[radius=\WFCDotRadius];
  \node[wfc/label,anchor=north] at (#1,-0.14) {$#2$};
}
\mathord{\vcenter{\hbox{%
\begin{tikzpicture}[x=\WFCUnit,y=\WFCUnit,line cap=round,line join=round]
  \begin{scope}
    \node[wfc/formula] at (0,\WFCFormulaY)
      {$\displaystyle\WFCOriginalFormula$};
    \WFCOuterLegs
    \WFCMiddleLeg
    \draw[wfc/internal] (-\WFCVertexStep,0)--(0,0)--(\WFCVertexStep,0);
    \WFCBoundary
    \WFCVertex{-\WFCVertexStep}{1}
    \WFCVertex{0}{2}
    \WFCVertex{\WFCVertexStep}{3}
  \end{scope}
  \begin{scope}[shift={({2*\WFCColumnStep},0)}]
    \node[wfc/formula] at (0,\WFCFormulaY)
      {$\displaystyle\WFCVertexCutFormula$};
    \WFCOuterLegs
    \draw[wfc/external] (-\WFCVertexStep,0)
      .. controls ({-0.55*\WFCVertexStep},{0.32*\WFCBoundaryHeight})
              and (-0.20,{0.70*\WFCBoundaryHeight})
      .. (0,\WFCBoundaryHeight);
    \draw[wfc/external] (0,\WFCBoundaryHeight)
      .. controls (0.27,{0.44*\WFCBoundaryHeight})
              and ({0.64*\WFCVertexStep},0.02)
      .. (\WFCVertexStep,0);
    \draw[wfc/boundary-leg] (0,\WFCBoundaryHeight)
      .. controls ({0.35*\WFCBoundaryLegSpan},{\WFCBoundaryHeight-\WFCBoundaryLegDepth})
              and ({0.75*\WFCBoundaryLegSpan},{\WFCBoundaryHeight-\WFCBoundaryLegDepth})
      .. (\WFCBoundaryLegSpan,\WFCBoundaryHeight);
    \WFCBoundary
    \fill[black!\WFCGrayTone] (0,\WFCBoundaryHeight) circle[radius=\WFCDotRadius];
    \node[wfc/label,text=black!\WFCGrayTone,anchor=north]
      at (0,{\WFCBoundaryHeight-0.24}) {$2$};
    \WFCVertex{-\WFCVertexStep}{1}
    \WFCVertex{\WFCVertexStep}{3}
  \end{scope}
  \begin{scope}[shift={(\WFCColumnStep,0)}]
    \node[wfc/formula] at (0,\WFCFormulaY)
      {$\displaystyle\WFCEdgeCutFormula$};
    \WFCOuterLegs
    \WFCMiddleLeg
    \draw[wfc/external] (-\WFCVertexStep,0)
      -- ({\WFCCutCenterX-\WFCCutHalfGap},\WFCBoundaryHeight);
    \draw[wfc/external] (0,0)
      -- ({\WFCCutCenterX+\WFCCutHalfGap},\WFCBoundaryHeight);
    \draw[wfc/internal] (0,0)--(\WFCVertexStep,0);
    \WFCBoundary
    \WFCVertex{-\WFCVertexStep}{1}
    \WFCVertex{0}{2}
    \WFCVertex{\WFCVertexStep}{3}
  \end{scope}
\end{tikzpicture}%
}}}
\endgroup
%
$}
 \caption{Graphical illustrations for $\Psi_G^{\partial}$,
 $\Psi_{G\setminus e}^{\partial}$, and
 $\Psi_{G\setminus v}^{\partial}$.
 In the middle panel, the internal edge is cut into two half-legs
 attached to the boundary; a tick marks this cut in the graphical
 subscript. In the right panel, the gray vertex has been moved to
 the boundary.}
 \label{fig:wavefunction-cuts}
\end{figure}

When the wavefunction recursion~\eqref{eq:combinatorial-wavefunction-recursion} is derived by integration by parts,
time derivatives can make the boundary terms nonzero.
These contributions must be retained when reorganizing the Born
expansion.
In flat spacetime with time-independent couplings,
$\mathcal N_V^\partial$ and $\mathcal N_E^\partial$ have
time-independent coefficients and therefore commute with the
total vertex-time derivative $\partial_{t}$. Applying this derivative to the
complete integrand and integrating by parts gives
\begin{equation}
 \begin{aligned}
 &\int_{\Rpos^{|V|}}\dd\mathbf t\,
 \mathcal N_V^\partial\Biggl[
 \left(\prod_{i\in\mathrm{ext}}K_{k_i}(t_{v(i)})\right)\\
 &\quad\times\mathcal N_E^\partial\Biggl[
 \Bigl(\sum_{v\in V}\partial_{t_v}\Bigr)
 \prod_{e=uw\in E}G_{BB}(t_u,t_w;y_e)\Biggr]\Biggr]\\
 &\qquad=X_G\Psi_G^\partial-\sum_{v\in V}\Psi_{G\setminus v}^\partial.
 \end{aligned}
 \label{eq:derivative-ibp}
\end{equation}
Differentiating the external modes gives $-X_G\Psi_G^\partial$,
while the displayed derivative acts on the scalar bulk-to-bulk
kernels, producing edge cuts through
$(\partial_t+\partial_{t'})G_{BB}=K_y(t)K_y(t')$.
Here $G\setminus e$ denotes cutting $e$ into two boundary half-legs,
whose modes remain within the scope of the inherited vertex and
propagator operators.
The vertex term $\Psi_{G\setminus v}^\partial$ is the boundary
trace at $t_v=0$ after all operators have acted; the contribution
at infinity vanishes in the convergence domain.
These edge cuts and boundary traces are illustrated in
Fig.~\ref{fig:wavefunction-cuts} and give the ordinary recursion
in the first line of Eq.~\eqref{eq:derivative-wavefunction-recursion}.

The standard Born expansion~\eqref{eq:ordinary-born} retains the
scalar cut factor $1/(2y_e)$. Inserting the ordinary recursion pairs
each non-bridge cut with the sector in which that edge was already
cut. The paired terms have identical inherited half-leg modes and
operator actions, so the scalar cancellation applies. The surviving
recursion replaces $X_H$ by $X_{V(H)}$ while retaining the
vertex-boundary terms, defining the embedded coefficients below:
\begin{equation}
\left\{
 \begin{aligned}
 &\Psi_H^{\partial}
 =\frac1{X_H}\sum_{a\in E(H)\cup V(H)}
       \Psi_{H\setminus a}^{\partial},\\[1ex]
 &\Psi_H^{\partial,\mathrm{emb}}
 =\frac1{X_{V(H)}}\sum_{a\in E(H)\cup V(H)}
       \Psi_{H\setminus a}^{\partial,\mathrm{emb}}.
 \end{aligned}
 \right.
 \label{eq:derivative-wavefunction-recursion}
\end{equation}
Here $E(H)$ and $V(H)$ denote the bulk-to-bulk edges and bulk
vertices of $H$. The ambient graph $G$, site energies $p_v$, and
original leg labels remain fixed throughout.
Both recursions apply separately to each bulk component and
terminate when no bulk-to-bulk propagators remain. At this stage,
$\Psi_H^{\partial,\mathrm{emb}}=\Psi_H^\partial$, evaluated with
the inherited operators, contractions, and boundary data, with
all contact terms retained. For $\operatorname{Tr}(\phi^3)$ theory,
the coefficients factorize over disconnected bulk components,
as in Fig.~\ref{fig:born-comparison}.

Start with any initial Born cut $I\subseteq E$. Recursively reduce
the bulk components containing cycles by cutting a bridge or moving
a bulk vertex to the boundary. Once a component is a tree, including
a singleton, collect its remaining recursion into an embedded
coefficient. The explicit reductions therefore end at a forest.
Let $\eta$ record this history: the initial cut $I_\eta$ and the
subsequent steps $(H,a)$, where $H$ is the connected bulk component
immediately before the operation and $a$ is the chosen bridge or
bulk vertex. Steps inside the terminal embedded coefficients are
not recorded in $\eta$. Summing over these histories gives the
simplified Born rule
\begin{equation}
 \begin{aligned}
 \braket{G}^{\partial}
 ={}&\sum_{\eta}
 2^{|\pi_0(G\setminus I_\eta)|-|I_\eta|}
 \left(\prod_{e\in I_\eta}\frac1{y_e}\right)\\[-2pt]
 &\quad\times\left[
 \left(\prod_{(H,a)\in\eta}
       \frac1{X_{V(H)}}\right)
 \Psi_{G\setminus\eta}^{\partial,\mathrm{emb}}\right].
 \end{aligned}
 \label{eq:derivative-revised-born}
\end{equation}
The factors on the first line give the weight of the initial Born
cut. Each subsequent bridge cut or vertex-boundary reduction adds
only the factor $1/X_{V(H)}$, leaving previously generated
denominators unchanged. If $G\setminus I_\eta$ is already a forest,
the product over $(H,a)$ is empty and equals $1$.
The subscript $G\setminus\eta$ records the final forest together
with its cut half-legs and boundary vertices. The brackets retain
the inherited contractions and state sums, including those between
disconnected bulk components. Histories with the same final graph
can have different intermediate denominators; histories differing
only in the order of independent reductions are counted once.
Appendix~\ref{app:born-rule} gives the proof.

For the bubble, the first line below contains the three forest-cut
contributions, while the second contains the additional terms from
moving either vertex to the boundary:
\begin{equation}
 \vcenter{\hbox{\resizebox{0.85\linewidth}{!}{$\displaystyle
\begingroup
\newcommand{\PBESmallStep}{3.8mm}
\newcommand{\PBESmallLineWidth}{0.50pt}
\newcommand{\PBESmallDotRadius}{1.00pt}
\newcommand{\PBELargeStep}{10mm}
\newcommand{\PBELargeLineWidth}{0.80pt}
\newcommand{\PBELargeDotRadius}{2.00pt}
\newcommand{\PBECorrelatorScale}{0.82}
\newcommand{\PBEPartialStyle}{\textstyle}
\newcommand{\PBEPartialScale}{0.85}
\newcommand{\PBEPartialXShift}{-0.20mm}
\newcommand{\PBEPartialYShift}{-0.70mm}
\newcommand{\PBEArcHeight}{0.42}
\newcommand{\PBECutHalfLength}{0.16}
\newcommand{\PBEGrayTone}{57}
\newcommand{\PBELabelFont}{\small}
\newcommand{\PBERowSpace}{0.6ex}
\newcommand{\PBETermSpace}{\,}
\newcommand{\PBEWaveSymbol}{\Psi}
\newcommand{\PBESuperscript}{\partial,\mathrm{emb}}
%
\newcommand{\PBESmallBubble}[6]{%
  \begin{tikzpicture}[
    x=\PBESmallStep,y=\PBESmallStep,baseline=-0.5ex,
    line width=\PBESmallLineWidth,line cap=round,line join=round]
    \path[use as bounding box]
      (-0.14,{-0.75*\PBEArcHeight-\PBECutHalfLength-0.05})
      rectangle (1.14,{0.75*\PBEArcHeight+\PBECutHalfLength+0.05});
    \draw[draw=#1] (0,0)
      .. controls (0.25,\PBEArcHeight) and (0.75,\PBEArcHeight) .. (1,0);
    \draw[draw=#2] (0,0)
      .. controls (0.25,-\PBEArcHeight) and (0.75,-\PBEArcHeight) .. (1,0);
    \ifnum#5=1\relax
      \draw[draw=#1]
        (0.5,{0.75*\PBEArcHeight-\PBECutHalfLength})
        --(0.5,{0.75*\PBEArcHeight+\PBECutHalfLength});
    \fi
    \ifnum#6=1\relax
      \draw[draw=#2]
        (0.5,{-0.75*\PBEArcHeight-\PBECutHalfLength})
        --(0.5,{-0.75*\PBEArcHeight+\PBECutHalfLength});
    \fi
    \fill[#3] (0,0) circle[radius=\PBESmallDotRadius];
    \fill[#4] (1,0) circle[radius=\PBESmallDotRadius];
  \end{tikzpicture}%
}
\newcommand{\PBEUpperCut}{%
  \PBESmallBubble{black!\PBEGrayTone}{black}{black}{black}{1}{0}}
\newcommand{\PBELowerCut}{%
  \PBESmallBubble{black}{black!\PBEGrayTone}{black}{black}{0}{1}}
\newcommand{\PBEBothCut}{%
  \PBESmallBubble{black!\PBEGrayTone}{black!\PBEGrayTone}{black}{black}{1}{1}}
\newcommand{\PBELeftGray}{%
  \PBESmallBubble{black!\PBEGrayTone}{black!\PBEGrayTone}{black!\PBEGrayTone}{black}{0}{0}}
\newcommand{\PBERightGray}{%
  \PBESmallBubble{black!\PBEGrayTone}{black!\PBEGrayTone}{black}{black!\PBEGrayTone}{0}{0}}
\newcommand{\PBEWave}[1]{\PBEWaveSymbol^{\PBESuperscript}_{#1}}
%
\newcommand{\PBECorrelatorGraph}{%
  \begin{tikzpicture}[
    x=\PBELargeStep,y=\PBELargeStep,baseline=-0.5ex,
    line width=\PBELargeLineWidth,line cap=round,line join=round,
    every node/.style={font=\PBELabelFont,inner sep=1.4pt}]
    \draw (0,0)
      .. controls (0.25,\PBEArcHeight) and (0.75,\PBEArcHeight) .. (1,0);
    \draw (0,0)
      .. controls (0.25,-\PBEArcHeight) and (0.75,-\PBEArcHeight) .. (1,0);
    \fill (0,0) circle[radius=\PBELargeDotRadius];
    \fill (1,0) circle[radius=\PBELargeDotRadius];
    \node[left] at (0,0) {$x_1$};
    \node[right] at (1,0) {$x_2$};
    \node[above] at (0.5,{0.75*\PBEArcHeight}) {$y_1$};
    \node[below] at (0.5,{-0.75*\PBEArcHeight}) {$y_2$};
  \end{tikzpicture}%
}
%
\newcommand{\PBECorrelator}{%
  \begin{tikzpicture}[baseline=(pbe-body.base)]
    \node[inner sep=0pt,outer sep=0pt] (pbe-body)
      {\scalebox{\PBECorrelatorScale}{%
        $\displaystyle\left\langle\PBECorrelatorGraph\right\rangle$}};
    \node[anchor=west,inner sep=0pt,outer sep=0pt]
      at ([xshift=\PBEPartialXShift,yshift=\PBEPartialYShift]pbe-body.north east)
      {\scalebox{\PBEPartialScale}{$\PBEPartialStyle\partial$}};
  \end{tikzpicture}%
}
\begin{aligned}
  \PBECorrelator
  &=\frac{1}{y_1}\PBETermSpace\PBEWave{\PBEUpperCut}
    +\frac{1}{y_2}\PBETermSpace\PBEWave{\PBELowerCut}
    +\frac{1}{y_1y_2}\PBETermSpace\PBEWave{\PBEBothCut}
  \\[\PBERowSpace]
  &\quad+\frac{2}{x_1+x_2}\PBETermSpace\PBEWave{\PBELeftGray}
    +\frac{2}{x_1+x_2}\PBETermSpace\PBEWave{\PBERightGray}
\end{aligned}
\endgroup%
%
$}}}
 \label{eq:derivative-bubble-expansion}
\end{equation}
For example, applying Eq.~\eqref{eq:derivative-wavefunction-recursion}
to the embedded coefficient with the upper edge cut gives
\begin{equation}
 \vcenter{\hbox{\resizebox{0.8\linewidth}{!}{$\displaystyle
\begingroup
\newcommand{\PEMSmallStep}{3.8mm}
\newcommand{\PEMSmallLineWidth}{0.50pt}
\newcommand{\PEMSmallDotRadius}{1.00pt}
\newcommand{\PEMArcHeight}{0.42}
\newcommand{\PEMCutHalfLength}{0.16}
\newcommand{\PEMGrayTone}{57}
\newcommand{\PEMPrefactorSpace}{\,}
\newcommand{\PEMWaveSymbol}{\Psi}
\newcommand{\PEMSuperscript}{\partial,\mathrm{emb}}
%
\newcommand{\PEMSmallBubble}[6]{%
  \begin{tikzpicture}[
    x=\PEMSmallStep,y=\PEMSmallStep,baseline=-0.5ex,
    line width=\PEMSmallLineWidth,line cap=round,line join=round]
    \path[use as bounding box]
      (-0.14,{-0.75*\PEMArcHeight-\PEMCutHalfLength-0.05})
      rectangle (1.14,{0.75*\PEMArcHeight+\PEMCutHalfLength+0.05});
    \draw[draw=#1] (0,0)
      .. controls (0.25,\PEMArcHeight) and (0.75,\PEMArcHeight) .. (1,0);
    \draw[draw=#2] (0,0)
      .. controls (0.25,-\PEMArcHeight) and (0.75,-\PEMArcHeight) .. (1,0);
    \ifnum#5=1\relax
      \draw[draw=#1]
        (0.5,{0.75*\PEMArcHeight-\PEMCutHalfLength})
        --(0.5,{0.75*\PEMArcHeight+\PEMCutHalfLength});
    \fi
    \ifnum#6=1\relax
      \draw[draw=#2]
        (0.5,{-0.75*\PEMArcHeight-\PEMCutHalfLength})
        --(0.5,{-0.75*\PEMArcHeight+\PEMCutHalfLength});
    \fi
    \fill[#3] (0,0) circle[radius=\PEMSmallDotRadius];
    \fill[#4] (1,0) circle[radius=\PEMSmallDotRadius];
  \end{tikzpicture}%
}
\newcommand{\PEMUpperCut}{%
  \PEMSmallBubble{black!\PEMGrayTone}{black}{black}{black}{1}{0}}
\newcommand{\PEMBothCut}{%
  \PEMSmallBubble{black!\PEMGrayTone}{black!\PEMGrayTone}{black}{black}{1}{1}}
\newcommand{\PEMLeftGrayUpperCut}{%
  \PEMSmallBubble{black!\PEMGrayTone}{black!\PEMGrayTone}{black!\PEMGrayTone}{black}{1}{0}}
\newcommand{\PEMRightGrayUpperCut}{%
  \PEMSmallBubble{black!\PEMGrayTone}{black!\PEMGrayTone}{black}{black!\PEMGrayTone}{1}{0}}
\newcommand{\PEMWave}[2][\PEMSuperscript]{\PEMWaveSymbol^{#1}_{#2}}
\PEMWave{\PEMUpperCut}
=\frac{1}{x_1+x_2}\PEMPrefactorSpace
  \left(
    \PEMWave[\partial]{\PEMBothCut}
    +\PEMWave[\partial]{\PEMLeftGrayUpperCut}
    +\PEMWave[\partial]{\PEMRightGrayUpperCut}
  \right)
\endgroup%
%
$}}}
 \label{eq:derivative-bubble-embedded-expansion}
\end{equation}
No bulk-to-bulk propagators remain on the right, so the embedded
and ordinary coefficients coincide there. For nonderivative
scalars, the vertex-boundary reductions vanish, leaving only
forest cuts and recovering Eq.~\eqref{eq:revised-born}.
Appendix~\ref{app:born-rule} illustrates the full construction
through an explicit calculation of the Yang--Mills bubble.

Upon fully expanding the embedded coefficients, histories without
vertex-boundary reductions reproduce the forest-cut and
maximal-tubing denominator patterns of $\operatorname{Tr}(\phi^3)$
theory. Boundary reductions give subproducts of these patterns,
with some $X_S$, $p_v$, or $y_e$ pole factors omitted. Thus the
scalar pole compatibility conditions still constrain the recursive
denominators, but do not guarantee nonzero joint residues, which
depend on the inherited operators and boundary data.

\section{Born polytope}

For conformally coupled Tr$(\phi^3)$ theory in flat spacetime,
a polytope description of the correlator~\cite{Arkani-Hamed:2024jbp,Figueiredo:2025daa}
must capture both pole compatibility and relative weights.
For graphs containing cycles, the standard Born-rule coefficient
$2^{|\pi_0(G\setminus I)|-|I|}$ varies between cut sectors
(Fig.~\ref{fig:born-comparison}). These relative weights go beyond
pole combinatorics, so identifying compatible poles with intersecting
facets does not, by itself, reproduce the standard expansion:
canonical forms have fixed boundary normalizations~\cite{Arkani-Hamed:2017tmz}.
The simplified rule in Eq.~\eqref{eq:revised-born} resolves this
mismatch through its uniform weight $2^{|V|-|E|}$.
Its pole products define the vertices of a simple $|E|$-dimensional
polytope, which we call the Born polytope, $\Born(G)$.

\begin{figure}[!htb]
	\centering
	\resizebox{0.753083\linewidth}{!}{$\displaystyle\input{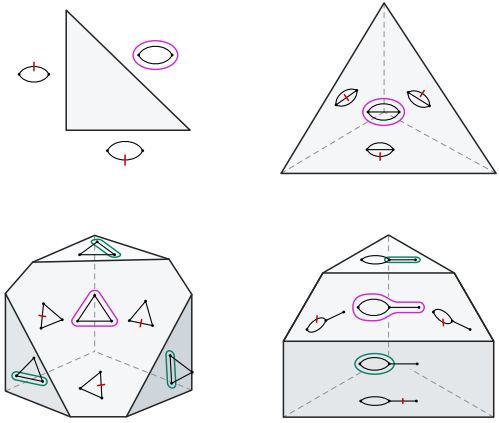}$}
	\caption{Born polytopes for a bubble,
		a three-edge watermelon, a triangle, and a tailed bubble. The small graphs
		label facets: a red cut marks an edge-energy pole $y_e$, a green tube
		marks a proper induced-subgraph pole $X_S$, and a magenta tube marks
		the total-energy pole $X_V$.}
	\label{fig:born-polytope}
\end{figure}

Let $\Conn(G)$ denote the vertex sets $S\subseteq V$ for which
$G[S]$ is connected and $|S|\geq2$.
The facets $\Facet(e)$ and $\Facet(S)$ are labeled by the physical
poles $y_e$ and $X_S$, respectively (Fig.~\ref{fig:born-polytope}).
For distinct facets, adjacency is characterized by
\begin{equation}
 \left\{
 \begin{aligned}
 \Facet(e){\sim}\Facet(f)
 &\Longleftrightarrow e\neq f,\\
 \Facet(S){\sim}\Facet(T)
 &\Longleftrightarrow \text{$S,T$ are nested or disjoint},\\
 \Facet(e){\sim}\Facet(S)
 &\Longleftrightarrow G[S]\setminus\{e\}\text{ is connected},
 \end{aligned}
 \right.
 \label{eq:facet-adjacency}
\end{equation}
where $\sim$ denotes adjacency along a codimension-two face.
Full facet intersections are specified by the forest cuts and
maximal tubings in the expanded rule~\eqref{eq:revised-born}:
each vertex corresponds to one pole product
(Appendix~\ref{app:minkowski}).
For trees, $\Born(G)$ recovers the graph correlahedron, or
$\mathcal G$-cubeahedron~\cite{Figueiredo:2025daa,tropicalstringyinin}.

\subsection{Canonical form}

Combinatorially, $\Born(G)$ is obtained by truncating the forest
polytope~\cite{Note1}.
For each $S\in\Conn(G)$, the face to be truncated has
vertices representing forests that contain a spanning tree of $G[S]$.
The truncation supplies the facet $X_S$, or moves its supporting
hyperplane if it is already present. Minkowski sums of Newton polytopes,
the convex hulls of monomial exponent vectors, realize this face
structure with the adjacency relations in Eq.~\eqref{eq:facet-adjacency},
as proved in Appendix~\ref{app:minkowski}.

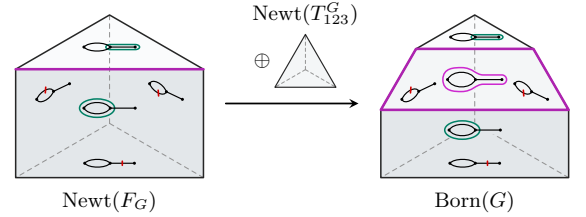
\begin{figure}[!htb]
 \centering
 \resizebox{0.85\linewidth}{!}{$\displaystyle
\begingroup
\newcommand{\TBUnit}{10mm}
\newcommand{\TBVertex}[4]{%
  \coordinate (#1) at
    ({0.707106781187*((#2)-(#3))},%
     {0.408248290464*(2*(#4)-(#2)-(#3))});}
\definecolor{TBShade}{HTML}{61788C}
\colorlet{TBFront}{TBShade!7}
\colorlet{TBTop}{TBShade!3}
\colorlet{TBLeft}{TBShade!18}
\definecolor{TBAccent}{HTML}{CD29CF}
\definecolor{TBGreen}{HTML}{008065}
\definecolor{TBRed}{HTML}{D40000}
\tikzset{
  tb/edge/.style={draw=black!85,line width=0.65pt},
  tb/hidden/.style={draw=black!42,line width=0.45pt,
    dash pattern=on 2pt off 1.7pt},
  tb/cut/.style={draw=TBAccent!85!black,line width=1.1pt},
  tb/label/.style={font=\small,inner sep=1pt},
  tb/graph/.style={draw=black,line width=0.50pt},
  tb/tube/.style={draw=TBGreen,line width=0.65pt},
  tb/alltube/.style={draw=TBAccent,line width=0.65pt},
  tb/graphcut/.style={draw=TBRed,line width=0.70pt},
  tb/dot/.style={circle,fill=black,draw=none,inner sep=0pt,
    minimum size=1.30pt}
}
\newcommand{\TBFacet}[7]{%
  \begin{scope}[shift={(#1,#2)},#7,scale=#3,shift={(-.285,0)}]
    \fill[#6] (0,0) ellipse[x radius=.43,y radius=.28];
    \fill[#6,rounded corners=1pt] (.20,-.13) rectangle (.99,.13);
    \ifnum#4=1
      \draw[tb/alltube]
        (-.40,0)
        .. controls (-.40,.2575) and (-.15,.2575) .. (0,.2575)
        .. controls (.25,.2575) and (.30,.10) .. (.47,.10)
        --(.87,.10) arc[start angle=90,end angle=-90,radius=.10]
        --(.47,-.10)
        .. controls (.30,-.10) and (.25,-.2575) .. (0,-.2575)
        .. controls (-.15,-.2575) and (-.40,-.2575) .. (-.40,0)
        --cycle;
    \fi
    \ifnum#4=2
      \draw[tb/tube] (0,0) ellipse[x radius=.40,y radius=.2575];
    \fi
    \ifnum#4=3
      \draw[tb/tube] (.30,-.10)--(.87,-.10)
        arc[start angle=-90,end angle=90,radius=.10]
        --(.30,.10) arc[start angle=90,end angle=270,radius=.10]
        --cycle;
    \fi
    \draw[tb/graph] (-.30,0)
      .. controls (-.20,.21) and (.20,.21) .. (.30,0);
    \draw[tb/graph] (-.30,0)
      .. controls (-.20,-.21) and (.20,-.21) .. (.30,0);
    \draw[tb/graph] (.30,0)--(.87,0);
    \foreach \x in {-.30,.30,.87} \node[tb/dot] at (\x,0) {};
    \ifnum#5=1 \draw[tb/graphcut] (0,.0725)--(0,.2425);\fi
    \ifnum#5=2 \draw[tb/graphcut] (0,-.2425)--(0,-.0725);\fi
    \ifnum#5=3 \draw[tb/graphcut] (.585,-.085)--(.585,.085);\fi
  \end{scope}%
}
\mathord{\vcenter{\hbox{%
\begin{tikzpicture}[x=\TBUnit,y=\TBUnit,line cap=round,line join=round]
  \begin{scope}
    \TBVertex{A}{0}{0}{0}
    \TBVertex{B}{2}{0}{0}
    \TBVertex{C}{0}{2}{0}
    \TBVertex{D}{0}{0}{2}
    \TBVertex{E}{2}{0}{2}
    \TBVertex{F}{0}{2}{2}
    \fill[TBLeft] (B)--(C)--(F)--(E)--cycle;
    \fill[TBTop] (D)--(E)--(F)--cycle;
    \draw[tb/hidden] (B)--(A)--(C) (A)--(D);
    \draw[tb/edge] (B)--(C)--(F)--(D)--(E)--cycle;
    \draw[tb/cut] (E)--(F);
    \TBFacet{0}{.23}{.66}{2}{0}{TBLeft}{cm={1.000000,0.000000,-0.000000,0.816497,(0,0)}}
    \TBFacet{0}{1.15}{.66}{3}{0}{TBTop}{cm={1.000000,0.000000,-0.000000,0.577350,(0,0)}}
    \TBFacet{-.85}{.48}{.60}{0}{1}{TBLeft}{cm={0.704416,0.335532,0.061628,0.848971,(0,0)}}
    \TBFacet{.85}{.48}{.60}{0}{2}{TBLeft}{cm={0.704416,-0.335532,-0.061628,0.848971,(0,0)}}
    \TBFacet{0}{-.60}{.64}{0}{3}{TBLeft}{cm={1.000000,0.000000,-0.000000,0.577350,(0,0)}}
    \node[tb/label] at (0,-1.18) {$\Newt(F_G)$};
  \end{scope}

  \draw[->,>=stealth,line width=0.75pt] (1.75,0.30)--(3.75,0.30);
  \node[tb/label] at (2.95,1.65) {$\Newt(T^G_{123})$};
  \node[tb/label] at (2.30,0.94) {$\oplus$};
  \begin{scope}[shift={(2.95,0.81)},scale=.65]
    \TBVertex{O}{0}{0}{0}
    \TBVertex{U}{1}{0}{0}
    \TBVertex{V}{0}{1}{0}
    \TBVertex{W}{0}{0}{1}
    \fill[TBFront] (U)--(V)--(W)--cycle;
    \draw[tb/hidden,line width=.40pt,
      dash pattern=on 1.4pt off 1.2pt] (O)--(U) (O)--(V) (O)--(W);
    \draw[tb/edge,line width=.50pt] (U)--(V)--(W)--cycle;
  \end{scope}

  \begin{scope}[shift={(5.50,0)}]
    \TBVertex{A}{0}{0}{0}
    \TBVertex{B}{2}{0}{0}
    \TBVertex{C}{0}{2}{0}
    \TBVertex{D}{0}{0}{2}
    \TBVertex{P}{2}{0}{1.25}
    \TBVertex{Q}{0}{2}{1.25}
    \TBVertex{R}{1.25}{0}{2}
    \TBVertex{S}{0}{1.25}{2}
    \fill[TBLeft] (B)--(C)--(Q)--(P)--cycle;
    \fill[TBTop] (D)--(R)--(S)--cycle;
    \fill[TBFront] (P)--(Q)--(S)--(R)--cycle;
    \draw[tb/hidden] (B)--(A)--(C) (A)--(D);
    \draw[tb/edge] (B)--(C)--(Q)--(S)--(D)--(R)--(P)--cycle;
    \draw[tb/cut] (P)--(Q)--(S)--(R)--cycle;
    \TBFacet{0}{1.30}{.62}{3}{0}{TBTop}{cm={1.000000,0.000000,-0.000000,0.577350,(0,0)}}
    \TBFacet{0}{.66}{.70}{1}{0}{TBFront}{cm={1.000000,0.000000,-0.000000,1.000000,(0,0)}}
    \TBFacet{-.85}{.48}{.60}{0}{1}{TBFront}{cm={0.704416,0.335532,0.061628,0.848971,(0,0)}}
    \TBFacet{.85}{.48}{.60}{0}{2}{TBFront}{cm={0.704416,-0.335532,-0.061628,0.848971,(0,0)}}
    \TBFacet{0}{-.10}{.66}{2}{0}{TBLeft}{cm={1.000000,0.000000,-0.000000,0.816497,(0,0)}}
    \TBFacet{0}{-.60}{.64}{0}{3}{TBLeft}{cm={1.000000,0.000000,-0.000000,0.577350,(0,0)}}
    \node[tb/label] at (0,-1.18) {$\Born(G)$};
  \end{scope}
\end{tikzpicture}%
}}}
\endgroup$}
 \caption{Minkowski realization of the tailed bubble.
 The polynomials are $F_G=(1+z_1+z_2)(1+z_3)$ and
 $T^G_{123}=1+z_1+z_2+z_3$.
 Magenta marks the truncated edge and the resulting facet.}
 \label{fig:tailed-bubble-truncation}
\end{figure}

We illustrate the construction with the tailed bubble in
Fig.~\ref{fig:born-comparison}.
Its forest polytope $\Newt(F_G)$ already has the cut facets $y_1,y_2,y_3$ and tube facets $X_{12},X_{23}$.
Adding $\Newt(T^G_{123})$ truncates the edge where the two tube facets
meet, creating the remaining facet $X_{123}$, as shown in
Fig.~\ref{fig:tailed-bubble-truncation}.

To recover the correlator, we choose the Minkowski weights and
translate the polytope so that its facet supports equal the physical
energies. For this graph, the resulting realization is
\begin{equation}
	\begin{aligned}
		&\{-\mathbf y\}\oplus x_1\Newt(1+z_1+z_2)\\
		&\oplus x_3\Newt(1+z_3) \oplus p_2\Newt(1+z_1+z_2+z_3),
	\end{aligned}
	\label{eq:tailed-bubble-physical-realization}
\end{equation}
where $\mathbf y=(y_1,y_2,y_3)$ and the point summand
translates the polytope by $-\mathbf y$.
Its canonical function, given by the
normalized polar-dual volume~\cite{stringycanonicalforms}, reproduces
the normalized correlator in Eq.~\eqref{eq:tailed-bubble}.

Consequently, the same polynomials and weights define a stringy deformation of
this canonical function~\cite{stringycanonicalforms}:
\begin{equation}
	\begin{aligned}
		 \int_{\Rpos^3}&\prod_{i=1}^3
		\frac{\ap z_i^{\ap y_i}\dd z_i}{z_i}(1+z_1+z_2)^{-\ap x_1}\\
		&(1+z_3)^{-\ap x_3}
		(1+z_1+z_2+z_3)^{-\ap p_2}.
	\end{aligned}
	\label{eq:tailed-bubble-stringy-integral}
\end{equation}
Each polynomial enters with exponent $-\ap$ times its Minkowski
weight, while the translation supplies the monomial factors
$z_i^{\ap y_i}$.

Appendix~\ref{app:minkowski} gives the general realization and
continues its canonical function analytically when the physical
weights lie outside the convex realization region.
The stringy formula also admits parallel-edge reduction; see
Appendix~\ref{app:stringy-derivations} for its derivation and the
full normalization.

\section{Outlook}

We have established the cancellation of all double-boundary poles
for arbitrary graphs in flat spacetime, including in theories with
local derivative interactions. The resulting simplified Born rule
organizes the surviving physical poles while retaining the required
boundary and contact terms. For conformally coupled scalars with
nonderivative interactions, the
uniform weights lead to the Born polytope and a new stringy formula.

The physical interpretation of embedded wavefunction coefficients
remains open. Do they admit a standard half-time integral
representation, and which boundary prescription would implement
the reorganization?

The explicit pole organization provides a starting point for
studying factorization~\cite{Arkani-Hamed:2025mce}.
How do physical-pole residues factorize for general graphs
with derivative interactions, once boundary and contact
contributions are included? Together with pole compatibility,
these data may constrain the correlator.

Time-dependent vertex weights in cosmological backgrounds produce
additional bulk terms in the integration-by-parts
step~\eqref{eq:derivative-ibp}. Dressing representations
\cite{Chowdhury:2025ohm,Chowdhury:2026dwm,Hillman:2021differential}
therefore suggest a route to applying the simplified expansion in
de Sitter space. Can double-boundary-pole cancellations in the flat-space
input constrain the singularities of the dressed correlator,
including its branch points? How do boundary terms and
regularization enter this relation
\cite{AguiSalcedo:2023tree,Bhowmick:2025singularities}?

For scalars, the Born polytope gives pole compatibility a geometric
meaning through facet incidence. It remains to determine whether
these incidence data can be extended to a geometric representation
of derivative correlators that also captures their theory-dependent
coefficients and boundary contributions.

\section*{Acknowledgments}

We thank Song He, Xiang Li, and Yong-Xiang Su for helpful discussions
and collaboration on related work. We also thank Dongyu Yang for
helpful discussions.
OpenAI's GPT-5.6 Sol and GPT-6 Astra models, accessed through Codex,
were used under the author's direction to assist with derivations,
computational checks, code development, and manuscript preparation.
The author reviewed and verified all AI-assisted material and scientific
conclusions and takes full responsibility for the content. 
This work is supported in part by the Science Challenge Project
(No.~TZ2025012) and NSAF (No.~U2330401).

\bibliography{references}

\onecolumngrid
\appendix
\setcounter{secnumdepth}{2}
\numberwithin{equation}{section}
\begin{center}
 \large\bfseries Supplemental material
\end{center}

\section{Proofs and examples of the simplified Born rule}
\label{app:born-rule}

The graph constructions assume no self-loops and retain parallel
edges as distinct. For nonderivative scalars, a self-loop contributes
only its factor $1/(2y_e)$ to the full-time correlator and can be
extracted first, with all graph and energy labels then referring to
the remaining graph.

\subsection{Nonderivative scalar theories}
\label{app:born-proof}

We first prove the nonderivative scalar formula~\eqref{eq:revised-born}. 
The core of the proof is the local cancellation illustrated in
Fig.~\ref{fig:spurious-cancellation}. Let $H+e$ be obtained from a
connected subgraph $H$ by adding an edge between its vertices.
Since $X_H-X_{H+e}=2y_e$,
\begin{equation}
 \frac{2}{X_{H+e}X_H}+\frac{1}{y_eX_H}
 =\frac{1}{y_eX_{H+e}}.
 \label{eq:double-boundary-local-cancellation}
\end{equation}
The double-boundary pole $X_H$ cancels, while the edge pole $y_e$
and any common inner wavefunction factors remain. This allows the
same cancellation to be repeated inside the remaining wavefunctions.

\begin{figure}[!htb]
 \centering
 \resizebox{0.9\linewidth}{!}{$\displaystyle\input{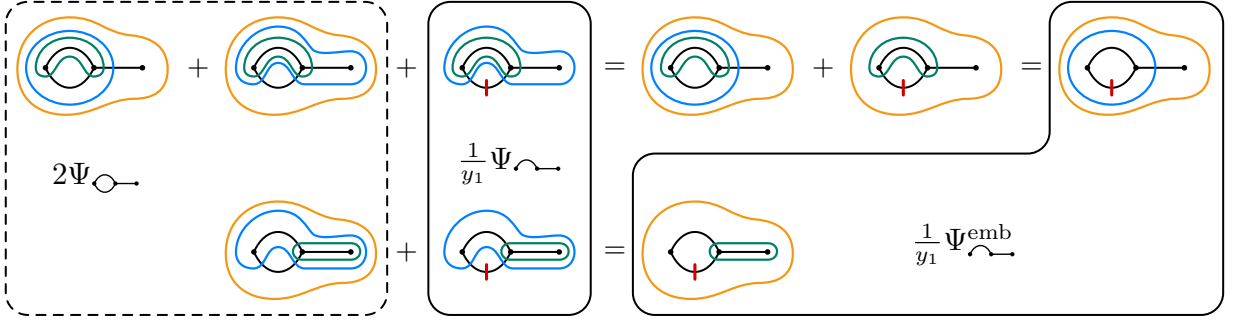}$}
 \caption{Double-boundary cancellations for the tailed bubble.
 The dashed box contains selected terms of $2\Psi_G$; the solid boxes
 contain complete tubing expansions before (middle) and after
 (right) reorganization. Common singleton factors are omitted.}
 \label{fig:tailed-bubble-reorganization}
\end{figure}

As a warm-up, consider the tailed bubble in
Fig.~\ref{fig:tailed-bubble-reorganization}. In the first row, the
outer double-boundary pole cancels first; the retained $1/y_1$ factor
participates in the subsequent cancellation of the inner pole. 
In each cancellation, the tubes share the same vertex set and differ
by the edge whose pole appears in the cut term.

We now organize this cancellation for a general graph using the
wavefunction recursion~\eqref{eq:combinatorial-wavefunction-recursion}.
For nonempty $S\subseteq V$, let $H$ run over connected spanning
subgraphs of $G[S]$. In the edge sums and products below, we use graphs
as shorthand for their edge sets. Define
\begin{equation}
 \mathcal W(S)
 :=\sum_H
    \left(\prod_{e\in G[S]\setminus H}\frac1{2y_e}\right)\Psi_H.
 \label{eq:connected-born-sector}
\end{equation}
Thus $\mathcal W(\{v\})=1/p_v$, and $\mathcal W(S)=0$ if $G[S]$ is
disconnected. A cut is uniquely specified by its component vertex sets
and a connected spanning subgraph on each set. Thus the flat-space
scalar specialization of Eq.~\eqref{eq:ordinary-born} becomes
\begin{equation}
 \braket{G}
 =\sum_{\Pi\in\operatorname{Part}(V)}
   \left(\prod_{e\in G\setminus\bigcup_{S\in\Pi}G[S]}\frac1{2y_e}\right)
   \prod_{S\in\Pi}2\mathcal W(S),
 \label{eq:born-partition-regrouping}
\end{equation}
where $\operatorname{Part}(V)$ denotes unordered set partitions, and
the edge product accounts for cuts between distinct blocks.

For $|S|\geq2$, multiply Eq.~\eqref{eq:connected-born-sector} by
$X_S=X_H-2\sum_{e\in G[S]\setminus H}y_e$ and use
Eq.~\eqref{eq:combinatorial-wavefunction-recursion}:
\begin{equation}
 X_S\mathcal W(S)
 =\sum_H
 \left(\prod_{f\in G[S]\setminus H}\frac1{2y_f}\right)
 \left(
  \sum_{e\in H}
  \prod_{K\in\pi_0(H\setminus\{e\})}\Psi_K
  -\sum_{e\in G[S]\setminus H}2y_e\Psi_H
 \right)\,.
 \label{eq:connected-born-sector-expansion}
\end{equation}
If $e$ is a non-bridge of $H$, then $H\setminus\{e\}$ is connected,
so the first inner sum in Eq.~\eqref{eq:connected-born-sector-expansion}
contributes $\Psi_{H\setminus\{e\}}$.
Its weighted contribution cancels the term in the second inner sum
for $H\setminus\{e\}$ with missing edge $e$, whose cut weight has an
extra factor $1/(2y_e)$:
\begin{equation}
 \left(\prod_{f\in G[S]\setminus H}\frac1{2y_f}\right)
 \left(\Psi_{H\setminus\{e\}}
       -\frac1{2y_e}2y_e\Psi_{H\setminus\{e\}}\right)=0.
 \label{eq:nonbridge-weight-identity}
\end{equation}
This is the local cancellation~\eqref{eq:double-boundary-local-cancellation}
organized after multiplying the connected sum by $X_S$, with the common
inner wavefunction left unexpanded.
This pairing exhausts the subtraction terms, since every missing edge
$e$ is a non-bridge of $H+e$. Thus only bridge deletions remain:
\begin{equation}
 X_S\mathcal W(S)
 =\sum_H
   \left(\prod_{f\in G[S]\setminus H}\frac1{2y_f}\right)
   \sum_{e\in\operatorname{Bri}(H)}\Psi_{H_L}\Psi_{H_R},
 \label{eq:surviving-bridge-sum}
\end{equation}
where $\operatorname{Bri}(H)$ is the set of bridges of $H$, and
$H_L,H_R$ are the two components of $H\setminus\{e\}$.
For each unordered vertex bipartition $S=A\sqcup B$ and joining edge
$e$, the component sums give $\mathcal W(A)\mathcal W(B)$, while every
other ambient edge between $A$ and $B$ must be cut. Thus
\begin{equation}
 \mathcal W(S)
 =\frac{1}{X_S}
  \sum_{\substack{S=A\sqcup B\\e\text{ joins }A\text{ and }B}}^{\mathrm{unordered}}
  \left(\prod_{\substack{f\ne e\\f\text{ joins }A\text{ and }B}}
    \frac1{2y_f}\right)
  \mathcal W(A)\mathcal W(B),\qquad |S|\geq2.
 \label{eq:connected-born-recursion}
\end{equation}
We iterate Eq.~\eqref{eq:connected-born-recursion} down to singleton
sets. In each term, the selected joining edges form a spanning tree
$T$ of $G[S]$, and we collect terms with the same $T$.
Each edge outside $T$ is cut once, when its endpoints are first
separated. For fixed $T$, each recursive bipartition deletes one edge
of the current subtree. After extracting the common cut factors, the
remaining sum obeys the embedded recursion~\eqref{eq:embedded-wavefunction}
with the same singleton initial values, and hence equals
$\Psi_T^{\mathrm{emb}}$. Thus
\begin{equation}
 \mathcal W(S)
 =\sum_{T\in\ST(G[S])}
   \left(\prod_{e\in G[S]\setminus T}\frac1{2y_e}\right)
   \Psi_T^{\mathrm{emb}},
 \label{eq:connected-born-tree-sum}
\end{equation}
where $\ST(G[S])$ denotes the set of spanning trees of $G[S]$.
The sum over one tree per partition block is the sum over forest cuts.
Substitution into Eq.~\eqref{eq:born-partition-regrouping} gives the
simplified Born rule~\eqref{eq:revised-born}, with coefficient
$2^{|\pi_0(G\setminus I)|-|I|}=2^{|V|-|E|}$ for each forest cut $I$.

\subsection{Derivative theories}
\label{app:derivative-born-proof}

We derive Eq.~\eqref{eq:derivative-revised-born} from the flat-space
Born expansion~\eqref{eq:ordinary-born} and the inherited
recursions~\eqref{eq:derivative-wavefunction-recursion}.
With time-independent coefficients, the vertex and propagator
operators each commute with the total vertex-time derivative,
giving the integration-by-parts identity~\eqref{eq:derivative-ibp};
no commutation between the two operators is required.
In Eq.~\eqref{eq:derivative-ibp}, cutting an edge $e=uv$ replaces
its scalar propagator by $K_{y_e}(t_u)K_{y_e}(t_v)$ within both
inherited operators. Boundary terms are traces of the complete
integrand after both operators act, with all contact contributions
retained.

Keep the ambient graph $G$ and singleton energies $p_v$ fixed
throughout, including after a vertex moves to the boundary.
For a bulk vertex set $S$ with fixed inherited external and boundary
data, sum over connected spanning subgraphs $H$ of $G[S]$:
\begin{equation}
 \mathcal W^\partial(S)
 =\sum_H
 \left(\prod_{e\in G[S]\setminus H}\frac1{2y_e}\right)
 \Psi_H^\partial.
 \label{eq:derivative-connected-born-sum}
\end{equation}
Here $\Psi_H^\partial$ includes the inherited half-leg modes and
their operator actions; only the scalar cut normalizations
$1/(2y_e)$ are displayed outside it.
For $|S|\geq2$, multiplying by
$X_S=X_H-2\sum_{e\in G[S]\setminus H}y_e$
and using the ordinary recursion gives the same non-bridge
cancellation as Eq.~\eqref{eq:nonbridge-weight-identity}.
Cutting $e$ in $H+e$ produces precisely the inherited state of
$H$, with the same half-leg modes, vertex and propagator operators,
contractions, and state sums. Its contribution cancels the subtraction
term for the missing edge $e$, leaving
\begin{equation}
 X_S\mathcal W^\partial(S)
 =\sum_H
 \left(\prod_{f\in G[S]\setminus H}\frac1{2y_f}\right)
 \left[
  \sum_{e\in\operatorname{Bri}(H)}\Psi_{H\setminus e}^\partial
  +\sum_{v\in S}\Psi_{H\setminus v}^\partial
 \right].
 \label{eq:derivative-surviving-reductions}
\end{equation}
We claim that iterating this cancellation gives
\begin{equation}
 \mathcal W^\partial(S)
 =\sum_H
 \left(\prod_{e\in G[S]\setminus H}\frac1{2y_e}\right)
 \sum_{\eta_H}\left[
 \left(\prod_{(K,a)\in\eta_H}
       \frac1{X_{V(K)}}\right)
 \Psi_{H\setminus\eta_H}^{\partial,\mathrm{emb}}\right].
 \label{eq:derivative-connected-history-sum}
\end{equation}
Here $\eta_H$ records reductions starting from $H$; the
missing-edge cuts $G[S]\setminus H$ are already included in the
prefactor. In each recorded pair $(K,a)$, $K$ is the connected bulk
component immediately before the operation, and $a$ is a bridge
or bulk vertex of $K$. Only components containing cycles are
reduced; a tree contributes its embedded coefficient with no
further steps recorded. If $H$ is already a tree, the history sum
contains a single empty history, with denominator product $1$.
For a singleton, both sides have the same inherited initial
value. To prove the identity by induction on $|S|$, multiply its
right-hand side by $X_S$ and expose the first reduction.
If $H$ is already a tree, expand its terminal embedded coefficient
once using Eq.~\eqref{eq:derivative-wavefunction-recursion}.
Every edge of a tree is a bridge, so this gives the same two types
of terms as Eq.~\eqref{eq:derivative-surviving-reductions}.

For either reduction, fix the resulting component vertex sets
$S_i$ and the retained joining edges $J$. For a bridge deletion,
$J=\{e\}$; for moving $v$ to the boundary, $J$ consists of the
retained edges incident to $v$, with at least one joining each
component. The parent graph and its cut weight then decompose as
\begin{equation}
 \begin{gathered}
 H=J\sqcup\bigsqcup_i H_i,\qquad
 C=G[S]\setminus\left(J\sqcup\bigsqcup_iG[S_i]\right),\\
 \prod_{e\in G[S]\setminus H}\frac1{2y_e}
 =\left(\prod_{e\in C}\frac1{2y_e}\right)
  \prod_i\left(\prod_{e\in G[S_i]\setminus H_i}\frac1{2y_e}\right).
 \end{gathered}
 \label{eq:derivative-reduction-cut-weights}
\end{equation}
Here each $H_i$ runs over connected spanning subgraphs of $G[S_i]$,
and $C$ is the fixed set of cuts outside them. Since $|S_i|<|S|$,
the induction hypothesis applies to each component sum within the
fixed inherited operator actions and boundary contractions, with
all parent denominators retained. This proves
Eq.~\eqref{eq:derivative-connected-history-sum}.

The tree stopping rule collects the remaining recursion after
cancellation; it does not replace an ordinary coefficient before
that cancellation. Independent reductions on distinct components
are counted once, regardless of their interleaving.

Finally, use the partition form of the standard Born rule,
Eq.~\eqref{eq:born-partition-regrouping}, with
$2\operatorname{Re}\mathcal W^\partial$ in place of
$2\mathcal W$ for each partition block and the inherited state sums
understood. Substituting
Eq.~\eqref{eq:derivative-connected-history-sum}, the cuts between
partition blocks and the missing edges within them combine into
$I_\eta$. Their coefficient is
\begin{equation}
 2^{|\pi_0(G\setminus I_\eta)|}
 \prod_{e\in I_\eta}\frac1{2y_e}
 =
 2^{|\pi_0(G\setminus I_\eta)|-|I_\eta|}
 \prod_{e\in I_\eta}\frac1{y_e}.
 \label{eq:derivative-history-weight}
\end{equation}
Each subsequent reduction $(H,a)\in\eta$ contributes its parent
factor $1/X_{V(H)}$, and the terminal inherited forest gives
$\Psi_{G\setminus\eta}^{\partial,\mathrm{emb}}$.
This is Eq.~\eqref{eq:derivative-revised-born}.

For nonderivative conformal scalars, the vertex-boundary
contributions vanish. Bridge cuts cannot remove a cycle, so only
initial forest cuts survive. Their coefficient is
$2^{|\pi_0(G\setminus I)|-|I|}=2^{|V|-|E|}$, recovering
Eq.~\eqref{eq:revised-born}.

\noindent\textbf{Tailed-bubble example.}
For the tailed bubble, label the vertices from left to right by
$1,2,3$, with $e_3$ the tail. The subscript ``forest-cut'' denotes
the terms in Eq.~\eqref{eq:derivative-revised-born} whose initial
cuts already leave a forest. The remaining boundary terms are
\begin{equation}
 \vcenter{\hbox{\resizebox{0.8\linewidth}{!}{$\displaystyle
\begingroup
\newcommand{\TBBSmallStep}{3.8mm}
\newcommand{\TBBSmallLineWidth}{0.50pt}
\newcommand{\TBBSmallDotRadius}{1.00pt}
\newcommand{\TBBCorrelatorStep}{4.5mm}
\newcommand{\TBBArcHeight}{0.42}
\newcommand{\TBBCutHalfLength}{0.16}
\newcommand{\TBBGrayTone}{57}
\newcommand{\TBBPrefactorSpace}{\,}
\newcommand{\TBBRowSpace}{0.6ex}
\newcommand{\TBBWaveSymbol}{\Psi}
\newcommand{\TBBSuperscript}{\partial,\mathrm{emb}}
\newcommand{\TBBForestCutLabel}{\text{forest-cut}}
%
\newcommand{\TBBSmallGraph}[6]{%
  \begin{tikzpicture}[
    x=\TBBSmallStep,y=\TBBSmallStep,baseline=-0.5ex,
    line width=\TBBSmallLineWidth,line cap=round,line join=round]
    \path[use as bounding box]
      (-0.14,{-0.75*\TBBArcHeight-\TBBCutHalfLength-0.05})
      rectangle (2.14,{0.75*\TBBArcHeight+\TBBCutHalfLength+0.05});
    \draw[draw=#1] (0,0)
      .. controls (0.25,\TBBArcHeight) and (0.75,\TBBArcHeight) .. (1,0);
    \draw[draw=#1] (0,0)
      .. controls (0.25,-\TBBArcHeight) and (0.75,-\TBBArcHeight) .. (1,0);
    \draw[draw=#2] (1,0)--(2,0);
    \ifnum#6=1\relax
      \draw[draw=#2] (1.5,-\TBBCutHalfLength)--(1.5,\TBBCutHalfLength);
    \fi
    \fill[#3] (0,0) circle[radius=\TBBSmallDotRadius];
    \fill[#4] (1,0) circle[radius=\TBBSmallDotRadius];
    \fill[#5] (2,0) circle[radius=\TBBSmallDotRadius];
  \end{tikzpicture}%
}
%
\newcommand{\TBBGraphA}{%
  \TBBSmallGraph{black!\TBBGrayTone}{black}{black!\TBBGrayTone}{black}{black}{0}}
\newcommand{\TBBGraphB}{%
  \TBBSmallGraph{black!\TBBGrayTone}{black!\TBBGrayTone}{black}{black!\TBBGrayTone}{black}{0}}
\newcommand{\TBBGraphC}{%
  \TBBSmallGraph{black!\TBBGrayTone}{black!\TBBGrayTone}{black}{black!\TBBGrayTone}{black!\TBBGrayTone}{0}}
\newcommand{\TBBGraphD}{%
  \TBBSmallGraph{black!\TBBGrayTone}{black!\TBBGrayTone}{black!\TBBGrayTone}{black}{black!\TBBGrayTone}{0}}
\newcommand{\TBBGraphE}{%
  \TBBSmallGraph{black!\TBBGrayTone}{black!\TBBGrayTone}{black}{black!\TBBGrayTone}{black}{1}}
\newcommand{\TBBGraphF}{%
  \TBBSmallGraph{black!\TBBGrayTone}{black!\TBBGrayTone}{black!\TBBGrayTone}{black}{black}{1}}
\newcommand{\TBBWave}[1]{\TBBWaveSymbol^{\TBBSuperscript}_{#1}}
\newcommand{\TBBFullGraph}{%
  \begingroup
    \renewcommand{\TBBSmallStep}{\TBBCorrelatorStep}%
    \TBBSmallGraph{black}{black}{black}{black}{black}{0}%
  \endgroup
}
\newcommand{\TBBCorrelator}{\left\langle\TBBFullGraph\right\rangle^{\partial}}
\begin{aligned}
\TBBCorrelator-\TBBCorrelator_{\TBBForestCutLabel} ={} &\frac{2}{X_{123}}\TBBPrefactorSpace
  \left(\TBBWave{\TBBGraphA}+\TBBWave{\TBBGraphB}\right)
+\frac{2}{X_{123}X_{12}}\TBBPrefactorSpace
  \left(\TBBWave{\TBBGraphC}+\TBBWave{\TBBGraphD}\right)
\\[\TBBRowSpace]
&\mathord{+}\left(\frac{2}{y_3X_{12}}+\frac{2}{X_{123}X_{12}}\right)\TBBPrefactorSpace
  \left(\TBBWave{\TBBGraphE}+\TBBWave{\TBBGraphF}\right)
\end{aligned}
\endgroup%
%
$}}}
 \label{eq:tailed-bubble-boundary}
\end{equation}
The term with $e_3$ cut and $v_1$ on the boundary in the second
line receives two contributions:
\begin{align}
 I_\eta=\{e_3\},\quad
 \eta=\{(G[\{v_1,v_2\}],v_1)\}
 &\quad\longmapsto\quad
 \frac{2}{y_3X_{12}}
 \Psi_{(G\setminus e_3)\setminus v_1}^{\partial,\mathrm{emb}},
 \label{eq:tailed-initial-cut-history}\\
 I_\eta=\varnothing,\quad
 \eta=\{(G,e_3),(G[\{v_1,v_2\}],v_1)\}
 &\quad\longmapsto\quad
 \frac{2}{X_{123}X_{12}}
 \Psi_{(G\setminus e_3)\setminus v_1}^{\partial,\mathrm{emb}}.
 \label{eq:tailed-recursive-cut-history}
\end{align}
Here $G[\{v_1,v_2\}]$ denotes the bubble component after cutting
$e_3$, with its inherited boundary half-leg. The initial Born weights
are $2/y_3$ and $2$, respectively; only the second history contains
a recursive bridge cut, which supplies $1/X_{123}$.
Moving $v_1$ to the boundary supplies $1/X_{12}$ in both cases.
Both histories stop at the same forest without further expanding
its embedded coefficient, so their weights add to the coefficient
in the second line of Eq.~\eqref{eq:tailed-bubble-boundary}.

\subsection{The Yang--Mills bubble in flat spacetime and de Sitter}
\label{app:born-examples}

We evaluate the $SU(N_c)$ gluon bubble in Coulomb
gauge~\cite{Campagnari:2009equal}, leaving the spatial loop momentum
unintegrated. We strip off $g^2N_c\delta^{AB}$, the symmetry factor
$1/2$, and the external covariances $1/(2k)^2$.
The external energies are $x_a=x_b=k$, so that
$X=x_a+x_b=2k$ and $p_a=p_b=k+y_1+y_2$.

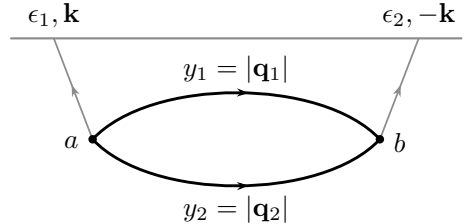
\begin{figure}[!htb]
 \centering
 \resizebox{0.35\linewidth}{!}{
\begingroup
\begin{tikzpicture}[
  x=1cm,y=1cm,line cap=round,line join=round,
  ym/boundary/.style={draw=black!45,line width=0.80pt},
  ym/external/.style={draw=black!45,line width=0.65pt},
  ym/internal/.style={draw=black,line width=1.06pt},
  ym/momentum/.style={postaction={decorate},
    decoration={markings,mark=at position 0.54 with
      {\arrow{Stealth[length=1.5mm,width=1.05mm]}}}},
  ym/label/.style={font=\normalsize,inner sep=1.2pt}
]
  \path[use as bounding box] (-3.0,-1.02) rectangle (3.0,1.78);
  \coordinate (a) at (-1.85,0);
  \coordinate (b) at ( 1.85,0);
  \draw[ym/external,ym/momentum] (a)--(-2.35,1.30);
  \draw[ym/external,ym/momentum] (b)--( 2.35,1.30);
  \draw[ym/boundary] (-2.90,1.30)--(2.90,1.30);
  \node[ym/label,anchor=south] at (-2.35,1.39)
    {$\epsilon_1,\mathbf k$};
  \node[ym/label,anchor=south] at ( 2.35,1.39)
    {$\epsilon_2,-\mathbf k$};
  \draw[ym/internal,ym/momentum]
    (a).. controls (-1.10,0.80) and (1.10,0.80)..(b);
  \draw[ym/internal,ym/momentum]
    (a).. controls (-1.10,-0.80) and (1.10,-0.80)..(b);
  \node[ym/label,anchor=south] at (0,0.68)
    {$y_1=|\mathbf q_1|$};
  \node[ym/label,anchor=north] at (0,-0.68)
    {$y_2=|\mathbf q_2|$};
  \fill[black] (a) circle[radius=1.55pt];
  \fill[black] (b) circle[radius=1.55pt];
  \node[ym/label,anchor=east] at (-1.98,-0.04) {$a$};
  \node[ym/label,anchor=west] at ( 1.98,-0.04) {$b$};
\end{tikzpicture}
\endgroup}
 \caption{The gluon bubble, with $y_i=|\mathbf q_i|$,
 $k=|\mathbf k|$, and
 $\epsilon_1\cdot\mathbf k=\epsilon_2\cdot\mathbf k=0$.}
 \label{fig:ym-bubble-kinematics}
\end{figure}

\noindent\textbf{The terminal expansion.}
Expanding every embedded coefficient in
Eqs.~\eqref{eq:derivative-bubble-expansion} and
\eqref{eq:derivative-bubble-embedded-expansion} gives
\begingroup
\newcommand{\ymBubbleSymbol}[6]{%
  \begin{tikzpicture}[
    x=3.8mm,y=3.8mm,baseline=-0.5ex,
    line width=0.50pt,line cap=round,line join=round]
    \path[use as bounding box] (-0.14,-0.525) rectangle (1.14,0.525);
    \draw[draw=#1] (0,0)
      .. controls (0.25,0.42) and (0.75,0.42) .. (1,0);
    \draw[draw=#2] (0,0)
      .. controls (0.25,-0.42) and (0.75,-0.42) .. (1,0);
    \ifnum#5=1\relax
      \draw[draw=#1] (0.5,0.155)--(0.5,0.475);
    \fi
    \ifnum#6=1\relax
      \draw[draw=#2] (0.5,-0.475)--(0.5,-0.155);
    \fi
    \fill[#3] (0,0) circle[radius=1.00pt];
    \fill[#4] (1,0) circle[radius=1.00pt];
  \end{tikzpicture}%
}
\newcommand{\ymBothCut}{%
  \ymBubbleSymbol{black!57}{black!57}{black}{black}{1}{1}}
\newcommand{\ymCutOneA}{%
  \ymBubbleSymbol{black!57}{black!57}{black!57}{black}{1}{0}}
\newcommand{\ymCutOneB}{%
  \ymBubbleSymbol{black!57}{black!57}{black}{black!57}{1}{0}}
\newcommand{\ymCutTwoA}{%
  \ymBubbleSymbol{black!57}{black!57}{black!57}{black}{0}{1}}
\newcommand{\ymCutTwoB}{%
  \ymBubbleSymbol{black!57}{black!57}{black}{black!57}{0}{1}}
\newcommand{\ymBoundaryA}{%
  \ymBubbleSymbol{black!57}{black!57}{black!57}{black}{0}{0}}
\newcommand{\ymBoundaryB}{%
  \ymBubbleSymbol{black!57}{black!57}{black}{black!57}{0}{0}}

\begin{equation}
\begin{aligned}
 \braket{C_2}^{\partial}
 ={}&\left[\frac1X\left(\frac1{y_1}+\frac1{y_2}\right)
       +\frac1{y_1y_2}\right]\Psi^{\partial}_{\ymBothCut}+\frac2X\left(
       \Psi^{\partial}_{\ymBoundaryA}
       +\Psi^{\partial}_{\ymBoundaryB}\right)\\[1ex]
 &+\frac1{Xy_1}\left(
       \Psi^{\partial}_{\ymCutOneA}
       +\Psi^{\partial}_{\ymCutOneB}\right)
  +\frac1{Xy_2}\left(
       \Psi^{\partial}_{\ymCutTwoA}
       +\Psi^{\partial}_{\ymCutTwoB}\right)
\end{aligned}

 \label{eq:ym-terminal-graphs}
\end{equation}
The upper and lower edges are $e_1,e_2$, and the left and right
vertices are $a,b$. A tick denotes a cut edge and a gray vertex
denotes a boundary restriction.

\noindent\textbf{The terminal integrals.}
To evaluate these coefficients, we use the edge operator $\Pi_e$
\begin{equation}
 \begin{aligned}
 (\Pi_e)_{\mu\nu}
 &=P_{\mu\nu}(\mathbf q_e)+\delta_{\mu0}\delta_{\nu0}
   \left(1-\frac{\partial_{t_a}^2+\partial_{t_b}^2}{2y_e^2}\right),
 &P_{ij}(\mathbf q)&=\delta_{ij}-\frac{q_iq_j}{|\mathbf q|^2}.
 \end{aligned}
 \label{eq:ym-edge-operator}
\end{equation}
Here $P_{ij}$ is the transverse projector, $0$ denotes Euclidean
time, and $i,j=1,2,3$ are spatial indices; we set
$P_{0\mu}=P_{\mu0}=0$.
Acting on $G_{BB}$, this operator produces the transverse and
instantaneous Coulomb propagators. Its $00$ component annihilates
$K_y(t_a)K_y(t_b)$, since $\partial_t^2K_y=y^2K_y$.
The same operator therefore gives only the transverse cut propagator
$P_{\mu\nu}K_y(t_a)K_y(t_b)/(2y)$.
We use the color-stripped cubic vertex
\begin{equation}
 \mathcal V_{\mu\alpha\beta}(d_0,d_1,d_2)
 =\delta_{\mu\alpha}(d_0-d_1)_\beta
  +\delta_{\alpha\beta}(d_1-d_2)_\mu
  +\delta_{\beta\mu}(d_2-d_0)_\alpha,
 \label{eq:ym-cubic-vertex}
\end{equation}
where each $d_r$ acts on its own leg, with spatial part $i\mathbf q_r$.
Writing this vertex operator at $a$ and $b$ as $\mathcal V_a$ and
$\mathcal V_b$, the graph operators for the bubble are
\begin{equation}
 \mathcal N_V^{\mathrm{YM}}
 =\epsilon_1^\mu\epsilon_2^\nu
   (\mathcal V_a)_{\mu\alpha\beta}
   (\mathcal V_b)_{\nu\gamma\delta},\qquad
 (\mathcal N_E^{\mathrm{YM}})_{\alpha\beta;\gamma\delta}
 =(\Pi_{e_1})_{\alpha\gamma}
   (\Pi_{e_2})_{\beta\delta}.
 \label{eq:ym-bubble-operators}
\end{equation}
Products involving the instantaneous Coulomb propagator,
$\delta(t_a-t_b)/y^2$, generate local double-contact terms formally
written as $\delta(t_a-t_b)^2$. In energy space, these terms contain
the scaleless factor $\int\dd q_0/(2\pi)\,1$, whose dimensionally
regulated continuation vanishes in split dimensional
regularization~\cite[Eq.~(18)]{Leibbrandt:1996split}.
We write $[\cdots]_{\mathrm{sub}}$ for the complete integrand with
this prescription applied after all vertex and propagator operators
have acted and before boundary restrictions. Single-contact and
boundary terms are retained.
The cancellation of energy divergences in the complete set of
Coulomb-gauge diagrams, including ghost contributions, is discussed
in Ref.~\cite[Sec.~III~D]{Niegawa:2006energy}.

Boundary restrictions use the symmetric endpoint trace
$\int_0^\infty\dd t\,\delta(t)f(t)=f(0)/2$.
With $P_i=P(\mathbf q_i)$, the fully cut graph gives
\begin{equation}
 \begin{aligned}
 \Psi^{\mathrm{YM}}_{\ymBothCut}
 &=\int_{\Rpos^2}\dd\mathbf t\,
 \Biggl[\mathcal N_V^{\mathrm{YM}}\Biggl[
 K_k(t_a)K_k(t_b)\,
 \mathcal N_E^{\mathrm{YM}}\left[
 \prod_{i=1}^2K_{y_i}(t_a)K_{y_i}(t_b)\right]\Biggr]\Biggr]_{\mathrm{sub}}\\
 &=\int_{\Rpos^2}\dd\mathbf t\,
 \epsilon_1^i\epsilon_2^j
 (\mathcal V_a)_{imn}(\mathcal V_b)_{jrs}
 P_1^{mr}P_2^{ns}\,e^{-p_at_a-p_bt_b}\\
 &=\frac{\epsilon_1^i\epsilon_2^j
 (\mathcal V_a)_{imn}(\mathcal V_b)_{jrs}
 P_1^{mr}P_2^{ns}}{p_ap_b},
 \end{aligned}
 \label{eq:ym-terminal-full-cut}
\end{equation}
For $e_2$ cut and $a$ on the boundary,
\begin{equation}
 \begin{aligned}
 \Psi^{\mathrm{YM}}_{\ymCutTwoA}
 &=\int_0^\infty\dd t_b\,
 \left.\Biggl[\mathcal N_V^{\mathrm{YM}}\Bigl[
 K_k(t_a)K_k(t_b)\,
 \mathcal N_E^{\mathrm{YM}}\bigl[
 G_{BB}(t_a,t_b;y_1)K_{y_2}(t_a)K_{y_2}(t_b)\bigr]\Bigr]\Biggr]_{\mathrm{sub}}
 \right|_{t_a=0}\\
 &=\int_0^\infty\dd t_b\,
 \frac{(\epsilon_1\cdot P_2\cdot\epsilon_2)(k-y_2)^2}{y_1^2}
       e^{-(k+y_2)t_b}\delta(t_b)\\
 &=\frac{(\epsilon_1\cdot P_2\cdot\epsilon_2)(k-y_2)^2}{2y_1^2}.
 \end{aligned}
 \label{eq:ym-terminal-cut-boundary-a}
\end{equation}
The corresponding restriction at $b$ gives
$\Psi^{\mathrm{YM}}_{\ymCutTwoB}=\Psi^{\mathrm{YM}}_{\ymCutTwoA}$.
The terms with $e_1$ cut follow by exchanging $1\leftrightarrow2$.
Only the instantaneous part of the uncut edge survives these
boundary restrictions. With neither edge cut and $a$ on the boundary,
\begin{equation}
 \begin{aligned}
 \Psi^{\mathrm{YM}}_{\ymBoundaryA}
 &=\int_0^\infty\dd t_b\,
 \left.\Biggl[\mathcal N_V^{\mathrm{YM}}\Biggl[
 K_k(t_a)K_k(t_b)\,
 \mathcal N_E^{\mathrm{YM}}\left[
 \prod_{i=1}^2G_{BB}(t_a,t_b;y_i)\right]\Biggr]\Biggr]_{\mathrm{sub}}
 \right|_{t_a=0}\\
 &=\int_0^\infty\dd t_b\,
 \sum_{i=1}^2\frac{(\epsilon_1\cdot P_i\cdot\epsilon_2)(k-y_i)}{y_{3-i}^2}
 e^{-(k+y_i)t_b}\delta(t_b)\\
 &=\frac12\sum_{i=1}^2
 \frac{(\epsilon_1\cdot P_i\cdot\epsilon_2)(k-y_i)}{y_{3-i}^2}.
 \end{aligned}
 \label{eq:ym-terminal-boundary-a}
\end{equation}
The corresponding restriction at $b$ gives
$\Psi^{\mathrm{YM}}_{\ymBoundaryB}=\Psi^{\mathrm{YM}}_{\ymBoundaryA}$.
Using $(k-y_i)^2/y_i+2(k-y_i)=(k^2-y_i^2)/y_i$ to combine
the boundary terms in Eq.~\eqref{eq:ym-terminal-graphs}, we obtain
\endgroup
\begin{equation}
 \braket{C_2}^{\mathrm{YM}}
 =\frac{X+y_1+y_2}{Xy_1y_2p_ap_b}
 \epsilon_1^i\epsilon_2^j
 (\mathcal V_a)_{imn}(\mathcal V_b)_{jrs}
 P_1^{mr}P_2^{ns}
 +\sum_{i=1}^2
 \frac{(\epsilon_1\cdot P_i\cdot\epsilon_2)(k^2-y_i^2)}
      {Xy_i y_{3-i}^2}.
 \label{eq:ym-bubble-result}
\end{equation}
The first term comes from two transverse internal propagators, and
the second from one transverse and one instantaneous Coulomb propagator.
After restoring the stripped factors and averaging over external
transverse polarizations, the two terms reproduce the bubble
contributions in Eqs.~(43) and (44) of
Ref.~\cite{Campagnari:2009equal}, up to loop-momentum relabeling.

In four dimensions, the conformal factor drops out of the classical
Yang--Mills action. With the same vacuum and boundary prescriptions
in conformal coordinates, the unintegrated result
\eqref{eq:ym-bubble-result} therefore also applies in $\mathrm{dS}_4$.

\section{Realization and canonical function of the Born polytope}
\label{app:minkowski}

We give a Minkowski-sum realization of the scalar Born polytope,
its facet structure, and the weights used in the stringy integral
of Appendix~\ref{app:stringy-derivations}.

\subsection{Minkowski summands and facet incidences}
\label{app:forest-refinement}

Let $\evec_e$ be the standard basis vectors of $\mathbb R^E$ and
$z_F=\prod_{e\in F}z_e$, with $z_\varnothing=1$.
Write $\cB(G)$ for the edge blocks of $G$~\cite{Note2}.
The forest polynomial is
\begin{equation}
 F_G(z):=
 \sum_{\substack{F\subseteq E\\(V,F)\text{ a forest}}}z_F
 =\prod_{B\in\cB(G)}F_B(z).
 \label{eq:forest-polynomial}
\end{equation}
Its Newton polytope is the forest polytope, or independence polytope
of the graphic matroid~\cite{Edmonds:1970polyhedra}.
For $S\in\Conn(G)$, let $A_G(S)$ be the union of the blocks containing
an edge of $G[S]$. A forest is $S$-separating if $S$ meets at least two
of its connected components, with connectivity tested in the whole
forest, including paths outside $S$. Define
\begin{equation}
 T_S^G(z)=
 \sum_{\substack{F\subseteq A_G(S)\\
                 F\text{ an }S\text{-separating forest}}}z_F.
 \label{eq:separating-polynomial}
\end{equation}
The polytope $\Newt(T_S^G)$ is the independence polytope of the
principal truncation along $E(G[S])$, restricted to the active blocks
$A_G(S)$~\cite[Prop.~3.1.5]{Backman:2024simplicial}.
These $0$--$1$ polytopes~\cite{Ziegler:2000zeroone} give the realization
\begin{equation}
 \Born(G)\simeq
 \bigoplus_{\mathclap{B\in\cB(G)}}\Newt(F_B)
 \oplus\bigoplus_{\mathclap{S\in\Conn(G)}}\Newt(T_S^G).
 \label{eq:minkowski}
\end{equation}
Here $\simeq$ denotes combinatorial equivalence. The tailed-bubble
instance is shown in Fig.~\ref{fig:tailed-bubble-truncation}.

For generic $w$, maximizing $\langle w,u\rangle$ over a Minkowski
sum amounts to maximizing it on each summand separately. The
directions for which a given vertex is a maximizer form its normal
cone, whose extreme rays are the outward normals of the facets
meeting there.
We prove Eq.~\eqref{eq:minkowski} by matching these cones to forest
cuts and maximal tubings. For a forest $F\subseteq E$ and the
non-singleton vertex sets $\mathcal L$ of a maximal tubing of $(V,F)$,
the cones are
\begin{equation}
 \mathcal C_{F,\mathcal L}
 =\operatorname{cone}\left(
 \{-\evec_e:e\notin F\}\cup
 \{\nu_T:T\in\mathcal L\}\right),
 \qquad \nu_T=\ind_{E(G[T])}.
 \label{eq:born-normal-cones}
\end{equation}
Here $\operatorname{cone}$ denotes the nonnegative span, and $\nu_T$
has entry $1$ on the edges of $G[T]$ and $0$ elsewhere.
To relate the separating summands to forests on all of $G$, restore
the inactive block factors:
\begin{equation}
 T_S^G(z)\prod_{\substack{B\in\cB(G)\\B\not\subseteq A_G(S)}}F_B(z)
 =\sum_{\substack{F'\subseteq E\text{ a forest}\\
                  F'\text{ separates }S}}z_{F'}.
 \label{eq:separating-inactive-factors}
\end{equation}
Choices in inactive blocks cannot change connectivity between
vertices of $S$. Products of polynomials correspond to Minkowski
sums of their Newton polytopes; the extra factors only add copies
of block summands already present, so they leave the normal cones
of the full sum unchanged.

For generic $w$, let $F$ maximize $\sum_{e\in F}w_e$ over forests.
Recursively split each non-singleton component with vertex set $T$
at its least-weight edge $e_T$, obtaining the tubing $\mathcal L$.
The maximizers on the right-hand side of
Eq.~\eqref{eq:separating-inactive-factors} are
\begin{equation}
 \underset{\substack{F'\subseteq E\text{ a forest}\\
                     F'\text{ separates }S}}{\operatorname{argmax}}
       \sum_{e\in F'}w_e
 =\begin{cases}
 F,&F\text{ separates }S,\\
 F\setminus\{e_T\},&\text{otherwise},
 \end{cases}
 \label{eq:principal-truncation-maximizer}
\end{equation}
where in the second case $T$ is the smallest tube containing $S$.
To see this, apply the greedy theorem for the principal
truncation~\cite[Sec.~40.1]{Schrijver:2003optimization}: process
positive-weight edges in decreasing order, rejecting an edge if
it creates a cycle or connects all vertices of $S$.
If $F$ connects $S$, the additional restriction first rejects
$e_T$, the least-weight edge of the minimal $F$-subtree spanning $S$.
An edge previously rejected for a cycle either still creates a cycle
or reconnects the two parts separated by $e_T$. The latter would
reconnect all of $S$, so no replacement edge is admitted.
Conversely, taking $S$ successively to be the vertex sets of the
current tree components identifies each deleted edge in
Eq.~\eqref{eq:principal-truncation-maximizer}, recovering every split.
Thus the summand maximizers and $(F,\mathcal L)$ determine each other.

To determine the entire normal cone, let $T^+$ be the smallest tube
strictly containing $T$, with $w_{e_{T^+}}:=0$ if no such tube exists.
For $e\notin F$, let $m_e$ be the minimum edge weight on the unique
$F$-path between its endpoints, or zero if they lie in different
components. The maximum-forest exchange inequalities give
$w_e<m_e$, while the recursive splits give
$w_{e_T}>w_{e_{T^+}}$. We then have
\begin{equation}
 w=\sum_{T\in\mathcal L}
       (w_{e_T}-w_{e_{T^+}})\nu_T
    -\sum_{e\notin F}(m_e-w_e)\evec_e.
 \label{eq:born-normal-cone-decomposition}
\end{equation}
For endpoints in one component, the smallest tube containing both
splits at the minimum-weight edge on their $F$-path. The first sum
therefore telescopes to $w_{e_U}$ on each forest edge $e_U$ and to
$m_e$ on each omitted edge.
Conversely, strictly positive coefficients of the generators in
Eq.~\eqref{eq:born-normal-cones} impose these same exchange and
split-ordering inequalities, and hence recover precisely $F$ and
$\mathcal L$. Taking closures proves the normal-cone formula,
including its boundary faces.

On the forest edges, $(\nu_T)_{e_U}=1$ precisely when $U\subseteq T$.
Ordering tubes and their split edges by inclusion therefore gives a
triangular matrix with unit diagonal. The cut coordinate vectors
supply the remaining unit axes, so the full set of normals is a
unimodular basis, with determinant $\pm1$. In particular, the Minkowski sum is simple, and
the facets at its vertex labeled by $(F,\mathcal L)$ are precisely
$y_e$ for $e\notin F$ and $X_T$ for $T\in\mathcal L$.
A collection of facets has a nonempty common intersection exactly
when its labels occur together in some forest--tubing term of the
simplified rule. This proves the combinatorial equivalence in
Eq.~\eqref{eq:minkowski}, including the adjacency relations in
Eq.~\eqref{eq:facet-adjacency}.
After facet matching, the normalized polar-dual volume therefore
reproduces the same pole products with unit coefficients
as $\tildd{\braket G}$~\cite{stringycanonicalforms}.

The truncation picture in the main text uses the forest-polytope faces
\begin{equation}
 \operatorname{face}_S=
 \left\{u\in\Newt(F_G):
   \sum_{e\in E(G[S])}u_e=|S|-1\right\},
 \qquad S\in\Conn(G).
 \label{eq:truncation-center}
\end{equation}
Their vertices are exactly the forests whose restriction to $S$
is a spanning tree of $G[S]$. A sufficiently shallow inward
displacement of this supporting hyperplane creates a facet with
normal $\nu_S$ and label $X_S$, or shifts its support if
$\operatorname{face}_S$ is already a facet.
The full incidence structure of the combined
Minkowski realization follows from Eq.~\eqref{eq:born-normal-cones};
individual separating summands need not be individual geometric cuts.

\noindent\textbf{Facet structure.}
Let $\Facet_{\mathrm{tot}}(H)$ denote the total-energy facet of
$\Born(H)$. The forest cuts and tubings give the following
combinatorial identifications:
\begin{equation}
\Facet(e)\simeq\Born(G\setminus\{e\}),\qquad
 \Facet(S)\simeq\Facet_{\mathrm{tot}}(G[S])\times\Born(G/S).
 \label{eq:facet-structure}
\end{equation}
For a disconnected graph, $\Born$ is understood as the Cartesian
product over its connected components. The quotient $G/S$ contracts
$S$ to one vertex, removes its internal edges, and retains all other
edges, including parallel edges.

The vertices of $\Facet(e)$ are exactly the forest--tubing pairs
with $e\notin F$, which are the pairs for $G\setminus\{e\}$ with the
same facet incidences. At a vertex of $\Facet(S)$, the tube $S$
ensures that $F|_S$ is a spanning tree of $G[S]$, giving the correspondence
\begin{equation}
 (F,\mathcal L)\longleftrightarrow
 \bigl((F|_S,\mathcal L|_S),(F/S,\mathcal L/S)\bigr).
 \label{eq:facet-restriction-contraction}
\end{equation}
Restriction keeps tubes contained in $S$; contraction keeps disjoint
tubes and contracts $S$ inside containing tubes, omitting the resulting
singleton. Internal cuts and proper subtubes belong to the first
factor, while external cuts and the remaining tubes belong to the
second. Replacing the contracted vertex by $S$ gives the unique
inverse and preserves these facet labels and their incidences,
proving Eq.~\eqref{eq:facet-structure}.
When $G[S]$ is a tree, $\Facet_{\mathrm{tot}}(G[S])$ is the graph
associahedron of its line graph~\cite[Thm.~2.6]{CarrDevadoss2006GraphAssociahedra}.

\subsection{Canonical function and physical weights}

For positive geometric weights $c_B,c_S$, take
\begin{equation}
 \mathsf Q_G
 =\bigoplus_{e\in E}y_e\Newt(z_e^{-1})
   \oplus\bigoplus_{B\in\cB(G)}c_B\Newt(F_B)
   \oplus\bigoplus_{S\in\Conn(G)}c_S\Newt(T_S^G).
 \label{eq:translated-minkowski}
\end{equation}
The point summands translate the polytope by $-\mathbf y$, where
$\mathbf y=(y_e)$. This preserves its face structure but fixes the
origin for the normalized polar-dual volume.
Write $h_{\mathsf P}(w)=\max_{u\in\mathsf P}\langle w,u\rangle$ and
$h_P=h_{\Newt(P)}$ for a polynomial $P$.
The cut supports are $h_{\mathsf Q_G}(-\evec_e)=y_e$.
For a connected $T$, its nonempty intersection with $V(B)$ induces
a connected subgraph of $B$: a simple path cannot leave a block
and return at a different vertex. A forest can therefore contain
at most $|T\cap V(B)|-1$ edges of $G[T]$ within $B$, and a spanning
tree of the intersection attains this bound. With
\begin{equation}
 q_B(T)=\max\{|T\cap V(B)|-1,0\},\qquad
 D_B=c_B+\sum_{\substack{S\in\Conn(G)\\B\subseteq A_G(S)}}c_S,
 \label{eq:combined-block-weights}
\end{equation}
the forest support is $h_{F_B}(\nu_T)=q_B(T)$, and $D_B$ collects
the weights multiplying this block contribution.
On $A_G(S)$, a forest maximizing the number of edges in $G[T]$
connects $S$ whenever $S\subseteq T$; separation then costs one edge.
Otherwise, a vertex of $S$ outside $T$ can be left isolated without
changing the support. Thus
\begin{equation}
 h_{T_S^G}(\nu_T)
 =\sum_{\substack{B\in\cB(G)\\B\subseteq A_G(S)}}q_B(T)
  -\ind_{S\subseteq T}.
 \label{eq:separating-support}
\end{equation}
Adding these supports, including the translation by $-\mathbf y$,
and matching $h_{\mathsf Q_G}(\nu_T)=X_T$ for independent tube
parameters $X_T$ gives
\begin{equation}
 \sum_{\substack{S\in\Conn(G)\\S\subseteq T}}c_S
 =\sum_BD_Bq_B(T)-X_T-\sum_{e\in E(G[T])}y_e
 =:f_X(T),\qquad T\in\Conn(G).
 \label{eq:geometric-mobius-f}
\end{equation}
Thus each facet fixes the sum of the weights for connected subsets
of $T$. Subtracting the weights of proper subsets recovers $c_T$;
iterating this subtraction is M\"obius inversion, also used to recover
Minkowski weights in Ref.~\cite[Prop.~2.3]{Ardila:2010matroid}.
Extend $f_X$ by zero to the empty set and singletons. On the inclusion
poset of connected vertex sets, including singletons and the empty
set, this gives
\cite[Sec.~3, Prop.~2]{Rota:1964mobius}
\begin{equation}
 c_S=\sum_{\substack{U\in\Conn(G)\\U\subseteq S}}
          \mu(U,S)f_X(U).
 \label{eq:mobius-inverse}
\end{equation}
For nonempty connected $U\subseteq S$, the minimal connected
enlargements of $U$ add one vertex adjacent to $U$. Their unions
cover $S$ only when every vertex of $S\setminus U$ is adjacent to
$U$, in which case all $|S|-|U|$ additions are required.
Inclusion--exclusion over these enlargements (the crosscut
formula~\cite[Sec.~6, Thm.~3]{Rota:1964mobius}) gives
\begin{equation}
 \mu(U,S)=
 \begin{cases}
 (-1)^{|S|-|U|},
 &\text{every vertex of }S\setminus U\text{ has a neighbor in }U,\\
 0,&\text{otherwise},
 \end{cases}
 \label{eq:mobius-connected}
\end{equation}
and $\mu(\varnothing,S)=-\sum_{\varnothing\neq U\subseteq S,\ G[U]\text{ connected}}\mu(U,S)$.
The block weights are
$c_B=D_B-\sum_{S:\,B\subseteq A_G(S)}c_S$, with the sum over $\Conn(G)$;
the $D_B$ parametrize the decomposition.

At the physical values
$X_T=\sum_{v\in T}p_v-2\sum_{e\in E(G[T])}y_e$,
write $(c_B,c_S,D_B)=(a_B,b_S,d_B)$. The data to be inverted become
\begin{equation}
 f_X(T)=\sum_Bd_Bq_B(T)-\sum_{v\in T}p_v
              +\sum_{e\in E(G[T])}y_e,
 \qquad T\in\Conn(G),
 \label{eq:physical-mobius-data}
\end{equation}
with $f_X$ still zero on the empty set and singletons.
The edge term assigns $y_e$ to every tube containing both endpoints
of $e$, so inversion places it only on that two-vertex set.
For a vertex $v\in S$, the identity
$-\sum_{U\in\Conn(G):\,v\in U\subseteq S}\mu(U,S)=\mu(\{v\},S)$
shows that its contribution is $\mu(\{v\},S)p_v$.
By Eq.~\eqref{eq:mobius-connected}, this is nonzero precisely when
$v$ is adjacent to every other vertex of $S$.

For the block contribution, apply the M\"obius identity to the
nonempty connected intersection $T\cap V(B)$. Separating the empty-set
contribution $1$ and each singleton contribution $-1$ gives
\begin{equation}
 q_B(T)=\sum_{\substack{U\in\Conn(G)\\U\subseteq T\cap V(B)}}
             \mu(\varnothing,U).
 \label{eq:block-rank-mobius}
\end{equation}
Both sides also vanish when $T\cap V(B)=\varnothing$.
Its inverse therefore contributes $d_B\mu(\varnothing,S)$ exactly
when $S\subseteq V(B)$. Combining the three contributions yields
\begin{equation}
 b_S=\ind_{|S|=2}\sum_{e\in E(G[S])}y_e
      +(-1)^{|S|-1}\sum_{v\in\Univ(G[S])}p_v
      +\mu(\varnothing,S)
        \sum_{\substack{B\in\cB(G)\\S\subseteq V(B)}}d_B,
 \label{eq:compact-weights}
\end{equation}
where $\Univ(G[S])$ is the set of vertices adjacent to every other
vertex of $G[S]$.

To recover $a_B$, delete the edges of $B$. Within the original
connected component containing $B$, each resulting component meets
$V(B)$ in exactly one vertex. A connected set in the original
component activates $B$ precisely when it is not contained in one
of these remaining components. Since
$f_X(T)=\sum_{S\in\Conn(G):\,S\subseteq T}b_S$, subtraction gives
\begin{equation}
 \begin{aligned}
 \sum_{\substack{S\in\Conn(G)\\B\subseteq A_G(S)}}b_S
 &=\sum_{K\in\pi_0(G)}f_X(V(K))
   -\sum_{K\in\pi_0(G\setminus B)}f_X(V(K))\\
 &=\sum_{e\in B}y_e-P_B+(|V(B)|-1)d_B,
 \end{aligned}
 \label{eq:active-block-weight-sum}
\end{equation}
where $P_B=\sum_{v\in V(B)\setminus\Cut(G)}p_v$ and $\Cut(G)$ is
the set of cut vertices. To obtain the second line, all other block
terms and all outside-edge contributions cancel. Vertex terms
cancel on nonsingleton components as well. The singleton components
created by the deletion have $f_X=0$ and are exactly the vertices
of $B$ that are not cut vertices. This leaves $-P_B$. Thus
\begin{equation}
 a_B=d_B-\sum_{\substack{S\in\Conn(G)\\B\subseteq A_G(S)}}b_S
       =P_B-\sum_{e\in B}y_e-(|V(B)|-2)d_B.
 \label{eq:auxiliary-block-weight}
\end{equation}
A convenient choice of the free block parameters is
\begin{equation}
 d_B=
 \begin{cases}
 p_u+p_v-\sum_{e\in B}y_e,&V(B)=\{u,v\},\\[3pt]
 \displaystyle\frac{P_B-\sum_{e\in B}y_e}{|V(B)|-2},&|V(B)|\geq3,
 \end{cases}
 \label{eq:block-choice}
\end{equation}
setting $a_B=0$ for $|V(B)|\geq3$ and the irrelevant weight of
$T_{\{u,v\}}^G=1$ to zero for two-vertex blocks.

Physical specialization can make some weights zero or negative and
leave the convex realization region. The canonical function, initially
the normalized polar-dual volume with the origin in the interior,
is rational in the facet parameters. Its analytic continuation to
the physical values gives $\tildd{\braket G}$.

\subsection{The triangle example}

\begin{figure}[htbp]
 \centering
 \resizebox{0.97\linewidth}{!}{$\input{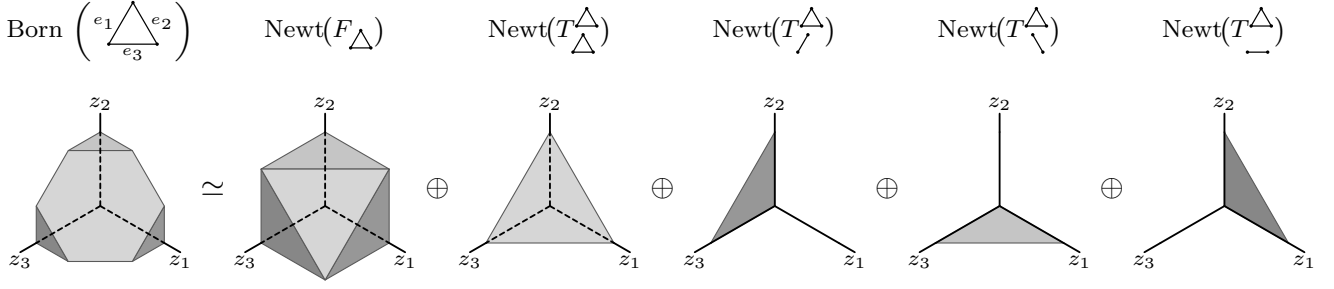}$}
 \newcommand{\BNCaptionGraph}[1]{\scalebox{0.65}{\BornMinkowskiGraph{#1}}}
 \caption[Minkowski realization of the Born polytope for $C_3$.]{%
  Minkowski realization of the Born polytope for $C_3$, with
  $F_{\BNCaptionGraph{0}}=1+z_1+z_2+z_3+z_1z_2+z_2z_3+z_3z_1$,
  $T^{\BNCaptionGraph{0}}_{\BNCaptionGraph{0}}=1+z_1+z_2+z_3$,
  and $T^{\BNCaptionGraph{0}}_{\BNCaptionGraph{1}}=1+z_2+z_3$.
  The other two polynomials follow by cyclic permutation.}
 \label{fig:minkowski}
\end{figure}

For $G=C_3$, take $e_1=(12)$, $e_2=(23)$, and $e_3=(31)$.
Figure~\ref{fig:minkowski} displays the forest and separating summands
in Eq.~\eqref{eq:minkowski}. Subscripts on $T$ below denote vertex sets.

There is a single block $E$. For independent facet parameters,
Eq.~\eqref{eq:geometric-mobius-f} gives
$f_X(123)=2D_E-X_{123}-y_1-y_2-y_3$.
Since $f_X$ vanishes on singletons, the inversion and block-weight
recovery reduce to
\begin{equation}
\begin{aligned}
 c_{12}&=f_X(12)=D_E-X_{12}-y_1
           \qquad\text{and cyclically},\\
 c_{123}&=f_X(123)-f_X(12)-f_X(23)-f_X(13)
        =X_{12}+X_{23}+X_{13}-X_{123}-D_E,\\
 c_E&=D_E-c_{12}-c_{23}-c_{13}-c_{123}
      =X_{123}+y_1+y_2+y_3-D_E.
 \end{aligned}
 \label{eq:triangle-mobius-inversion}
\end{equation}
Substituting the physical energies and
$D_E=d_E=p_1+p_2+p_3-y_1-y_2-y_3$ from
Eq.~\eqref{eq:block-choice} gives
\begin{equation}
 a_E=0,\qquad b_{123}=y_1+y_2+y_3,\qquad
 b_{12}=x_3,\quad b_{23}=x_1,\quad b_{13}=x_2.
 \label{eq:triangle-weights}
\end{equation}
Thus the translated realization is
\begin{equation}
 \mathsf Q_{C_3}
 =\{-\mathbf y\}
 \oplus(y_1+y_2+y_3)\Newt(T_{123}^{C_3})
 \oplus x_3\Newt(T_{12}^{C_3})
 \oplus x_1\Newt(T_{23}^{C_3})
 \oplus x_2\Newt(T_{13}^{C_3}),
 \label{eq:triangle-physical-realization}
\end{equation}
where $\mathbf y=(y_1,y_2,y_3)$. Although the forest summand has zero
weight, the remaining four summands retain the Born face structure
for $x_i,y_i>0$. The facet supports are
\begin{equation}
 h_{\mathsf Q_{C_3}}(-\evec_i)=y_i,\qquad
 h_{\mathsf Q_{C_3}}(\evec_1)=X_{12}\quad\text{and cyclically},\qquad
 h_{\mathsf Q_{C_3}}(\evec_1+\evec_2+\evec_3)=X_{123}.
 \label{eq:triangle-supports}
\end{equation}
The normalized polar-dual volume reproduces the ten pole products
of the simplified rule:
\begin{equation}
 \tildd{\braket{C_3}}
 =\frac1{y_1y_2y_3}
 +\sum_{\mathrm{cyc}}\frac1{X_{12}y_2y_3}
 +\frac1{X_{123}}\sum_{\mathrm{cyc}}\frac1{y_1}
   \left(\frac1{X_{23}}+\frac1{X_{31}}\right).
 \label{eq:triangle-canonical-function}
\end{equation}
Here the cyclic sums act on $1,2,3$. The corresponding stringy integral is
\begin{equation}
 \tildd{\braket{C_3}}^{\str}
 =\int_{\Rpos^3}\prod_{i=1}^3
       \frac{\ap z_i^{\ap y_i}\dd z_i}{z_i}
 \frac{\displaystyle\prod_{\mathrm{cyc}}
                  (1+z_2+z_3)^{-\ap x_3}}
 {(1+z_1+z_2+z_3)^{\ap(y_1+y_2+y_3)}}.
 \label{eq:triangle-stringy-integral}
\end{equation}
Its $\ap\to0$ limit is Eq.~\eqref{eq:triangle-canonical-function}.

\section{Stringy integral}
\label{app:stringy-derivations}

For the nonderivative scalar theory, the Minkowski weights obtained
by facet matching and M\"obius inversion in Appendix~\ref{app:minkowski}
give the stringy deformation of the stripped
correlator~\cite{stringycanonicalforms}:
\begin{equation}
 \tildd{\braket{G}}^{\str}=
 \int_{\Rpos^E}
       \prod_{e\in E}\frac{\ap z_e^{\ap y_e}\dd z_e}{z_e}
       \prod_{B\in\cB(G)}(F_B)^{-\ap a_B}
       \prod_{S\in\Conn(G)}(T_S^G)^{-\ap b_S},
 \label{eq:stringy-integral}
\end{equation}
with the weights in Eqs.~\eqref{eq:compact-weights},
\eqref{eq:auxiliary-block-weight}, and~\eqref{eq:block-choice}.
The translation by $-\mathbf y$ supplies $(z_e^{-1})^{-\ap y_e}=z_e^{\ap y_e}$ for each edge.
The tropical criterion of Ref.~\cite[Sec.~4.2, Claim~3]{stringycanonicalforms}
applies after physical specialization: the exponent
\begin{equation*}
 \sum_Ba_Bh_{F_B}(w)+\sum_{S\in\Conn(G)}b_Sh_{T_S^G}(w)
 -\langle\mathbf y,w\rangle
\end{equation*}
is linear on the cones in Eq.~\eqref{eq:born-normal-cones}, with ray
values $y_e$ and $X_T$. For $y_e,X_T>0$ it is positive away from the
origin, so Eq.~\eqref{eq:stringy-integral} converges for $\ap>0$ even
with zero or negative weights. As $\ap\to0$, the unimodular cone
contributions reproduce the forest-cut terms of the stripped simplified
rule, giving $\tildd{\braket G}$. Other kinematics are defined by
meromorphic continuation.

Restoring the vertex factors corresponds to a Cartesian product with
a $|V|$-dimensional hypercube, whose stringy integral gives the
product of beta factors below~\cite[Appendix~A]{tropicalstringyinin}:
\begin{equation}
 \braket{G}^{\str}
 :=\left[\frac{1}{2^{|E|}}\prod_{v\in V}
 \ap B(\ap p_v,\ap p_v)\right]\tildd{\braket{G}}^{\str},
 \label{eq:full-stringy-correlator}
\end{equation}
where $B$ is the Euler beta function. The prefactor reproduces
the normalization of Eq.~\eqref{eq:normalized-correlator} as $\ap\to0$.
An accompanying Mathematica notebook generates 1-skeleton diagrams of
Born polytopes and their stringy integrals.

\subsection{Parallel-edge reduction}
\label{app:parallel-reduction}

The scalar simplified Born rule makes parallel-edge reduction, or melonic
simplification~\cite{Donath:2024utn,Arkani-Hamed:2025mce}, manifest:
a forest retains at most one edge of a parallel bundle, and the embedded
wavefunctions are independent of which edge is retained. Grouping these
cuts replaces the bundle by a single edge carrying its total energy,
with an overall prefactor. The full stringy correlator satisfies the
corresponding finite-$\ap$ relation:
\begin{equation}
 \vcenter{\hbox{\resizebox{0.45\linewidth}{!}{$\displaystyle
\begingroup
\newcommand{\MELUnit}{6mm}
\newcommand{\MELBridge}{1.30}
\newcommand{\MELBlobRadius}{0.44}
\newcommand{\MELBlobFill}{black!15}
\newcommand{\MELLineWidth}{0.65pt}
\newcommand{\MELVertexRadius}{1.15pt}
\newcommand{\MELUpperHeight}{0.60}
\newcommand{\MELLowerHeight}{0.60}
\newcommand{\MELLabelFont}{\footnotesize}
\newcommand{\MELFactorSpace}{\,}
\newcommand{\MELStringyStyle}{\textstyle}
\newcommand{\MELStringyShift}{-8mu}
\newcommand{\MELStringyRaise}{-2.0ex}
\newcommand{\MELGammaScale}{0.95}
%
\newcommand{\MELGraph}[1]{%
  \mathord{\vcenter{\hbox{%
    \begin{tikzpicture}[
      x=\MELUnit,y=\MELUnit,
      line width=\MELLineWidth,line cap=round,line join=round,
      every node/.style={inner sep=1.2pt}]
      \coordinate (melL) at (0,0);
      \coordinate (melR) at (\MELBridge,0);
      \draw[fill=\MELBlobFill] (-\MELBlobRadius,0) circle[radius=\MELBlobRadius];
      \draw[fill=\MELBlobFill] ({\MELBridge+\MELBlobRadius},0) circle[radius=\MELBlobRadius];
      \node at (-\MELBlobRadius,0) {$L$};
      \node at ({\MELBridge+\MELBlobRadius},0) {$R$};
      \ifnum#1=1\relax
        \draw (melL)
          .. controls ({0.25*\MELBridge},\MELUpperHeight)
                  and ({0.75*\MELBridge},\MELUpperHeight) .. (melR);
        \draw (melL)
          .. controls ({0.25*\MELBridge},-\MELLowerHeight)
                  and ({0.75*\MELBridge},-\MELLowerHeight) .. (melR);
        \node[font=\MELLabelFont,above]
          at ({0.5*\MELBridge},{0.75*\MELUpperHeight}) {$y_1$};
        \node[font=\MELLabelFont,below]
          at ({0.5*\MELBridge},{-0.75*\MELLowerHeight}) {$y_r$};
        \foreach \melDotY in {-0.15,0,0.15}
          \fill ({0.5*\MELBridge},\melDotY) circle[radius=0.45pt];
      \else
        \draw (melL) -- (melR);
        \node[font=\MELLabelFont,above=1.5pt]
          at ({0.5*\MELBridge},0) {$\sum\limits_{i=1}^{r}y_i$};
      \fi
      \fill (melL) circle[radius=\MELVertexRadius];
      \fill (melR) circle[radius=\MELVertexRadius];
      \pgfresetboundingbox
      \path[use as bounding box]
        ({-2*\MELBlobRadius-0.06},-1.22)
        rectangle ({\MELBridge+2*\MELBlobRadius+0.06},1.22);
    \end{tikzpicture}%
  }}}%
}
\newcommand{\MELCorrelator}[1]{%
  {\left\langle\MELGraph{#1}\right\rangle}^{%
    \mkern\MELStringyShift
    \raisebox{\MELStringyRaise}[0pt][0pt]{\(\MELStringyStyle\mathrm{str}\)}%
  }%
}
\MELCorrelator{1}
=
\mathord{\vcenter{\hbox{\scalebox{\MELGammaScale}{\(\displaystyle
  \dfrac{\prod\limits_{i=1}^{r}\frac{\alpha'}{2}\,\Gamma\!\left(\alpha' y_i\right)}
        {\frac{\alpha'}{2}\,\Gamma\!\left(\alpha'\sum\limits_{i=1}^{r}y_i\right)}
\)}}}}
\MELFactorSpace\MELCorrelator{0}
\endgroup%
$}}}
 \label{eq:parallel-reduction}
\end{equation}

Let $G^{\mathrm{multi}}$ contain a bundle of $r$ parallel edges, and
let $G^{\mathrm{simple}}$ be obtained by replacing the bundle with one
edge of energy $\sum_i y_i$, keeping all $p_v$ fixed. The polynomials
$F_G$ and $T_S^G$ depend on $z_1,\ldots,z_r$ only through
$z_\parallel=\sum_i z_i$. In this variable, they coincide with the
corresponding polynomials for $G^{\mathrm{simple}}$.
Their exponents also agree: the physical weights depend on vertex
and block data unchanged by merging, and on the bundle energies
only through their sum.

Let $\mathcal I_G^{\mathrm{int}}$ denote the scalar integrand in
Eq.~\eqref{eq:stringy-integral}, with the measure
$\prod_e\ap\,\dd z_e/z_e$ omitted. Cancelling the common polynomial
factors gives
\begin{equation}
 \mathcal I_{G^{\mathrm{multi}}}^{\mathrm{int}}(z_1,\ldots,z_r,\ldots)
 =\frac{\prod_{i=1}^r z_i^{\ap y_i}}
        {z_\parallel^{\ap\sum_{i=1}^r y_i}}
  \mathcal I_{G^{\mathrm{simple}}}^{\mathrm{int}}(z_\parallel,\ldots),
 \label{eq:parallel-integrand}
\end{equation}
where the remaining edge variables are unchanged.
Setting $z_i=z_\parallel t_i$, with $\sum_i t_i=1$, turns the prefactor
into $\prod_i t_i^{\ap y_i}$. Including the full measure, the change
of variables and the Dirichlet integral give
\begin{equation}
\begin{aligned}
 \tildd{\braket{G^{\mathrm{multi}}}}^{\str}
 ={}&\int_{\Rpos^{E(G^{\mathrm{multi}})}}
       \left(\prod_{e\in E(G^{\mathrm{multi}})}
             \frac{\ap\,\dd z_e}{z_e}\right)
       \frac{\prod_{i=1}^r z_i^{\ap y_i}}
            {z_\parallel^{\ap\sum_i y_i}}
       \mathcal I_{G^{\mathrm{simple}}}^{\mathrm{int}}(z_\parallel,\ldots)
       \\
 ={}&(\ap)^{r-1}
       \left[\int_{\substack{t_i>0\\\sum_i t_i=1}}
          \prod_{i=1}^r t_i^{\ap y_i-1}\,
          \dd t_1\cdots\dd t_{r-1}\right]
       \int_{\Rpos^{E(G^{\mathrm{simple}})}}
          \left(\prod_{e\in E(G^{\mathrm{simple}})}
                \frac{\ap\,\dd z_e}{z_e}\right)
          \mathcal I_{G^{\mathrm{simple}}}^{\mathrm{int}}(z_\parallel,\ldots)
       \\
 ={}&(\ap)^{r-1}
       \frac{\prod_{i=1}^r\Gamma(\ap y_i)}{\Gamma(\ap\sum_i y_i)}
       \tildd{\braket{G^{\mathrm{simple}}}}^{\str}.
\end{aligned}
 \label{eq:dirichlet}
\end{equation}
Restoring the normalization in Eq.~\eqref{eq:full-stringy-correlator}
supplies $2^{1-r}$, while the vertex beta functions cancel, yielding
Eq.~\eqref{eq:parallel-reduction}. The reduction extends by meromorphic
continuation from the convergence domain.
As $\ap\to0$, the full prefactor tends to
$(\sum_i y_i)/(2^{r-1}\prod_i y_i)$, reproducing the parallel-edge
reduction in the field-theory limit.

\end{document}